# Optimizing Economic Complexity

Viktor Stojkoski[1,2] and César A. Hidalgo[1,3*†]

[1]Center for Collective Learning, CIAS, Corvinus University of Budapest, Budapest, Hungary
[2] Ss. Cyril and Methodius University in Skopje, Skopje, North Macedonia
[3]Center for Collective Learning, IAST, Toulouse School of Economics, Université de Toulouse Capitole, Toulouse, France

**Abstract**

Efforts to apply economic complexity to identify diversification opportunities often rely on diagrams comparing the relatedness and complexity of products, technologies, or industries. Yet, the use of these diagrams, is not based on empirical or theoretical evidence supporting some notion of optimality. Here, we introduce an optimization-based framework that identifies diversification opportunities by minimizing a cost function capturing the constraints imposed by an economy's pattern of specialization. We show that the resulting portfolios often differ from those implied by relatedness–complexity diagrams, providing a target-oriented optimization layer to the economic complexity toolkit.

## Introduction

To achieve sustainable economic growth economies must adapt to changes in markets and technologies [1,2]. Changes in economic structure, however, involve strategic considerations such as identifying diversification opportunities while facing the constraints imposed by an economy's existing productive structure [3–5].

During the last couple of decades, economic complexity has become a common method to explore questions of strategic diversification [6–30]. The use of these methods is supported by the notion that economic complexity predicts economic growth [6,9,13,14,31], and thus, connects changes in an economy's productive structure to its growth potential. Yet, the practical implementation of these ideas often relies on the use of *relatedness-complexity diagram*s: a graphical method introduced more than a decade ago [9] to visually inspect an economy's diversification opportunities [9,32–35].

* Corresponding author: cesar.hidalgo@tse-fr.eu
† We are supported by the European Union through the 101086712-LearnData-HORIZON-WIDERA-2022-TALENTS-01 financed by European Research Executive Agency (REA) (https://cordis.europa.eu/project/id/101086712), IAST funding from the French National Research Agency (ANR) under grant ANR-17-EURE-0010 (Investissements d'Avenir program), and the European Lighthouse of AI for Sustainability [grant number 101120237-HOR-IZON-CL4-2022-HUMAN-02].

In a relatedness-complexity diagram, economic activities, such as product exports, industries, or technologies, are presented in a two-way scatter plot (Figure 1) with relatedness in the x-axis and complexity on the y-axis. These diagrams are usually divided into quadrants to help facilitate their interpretation.

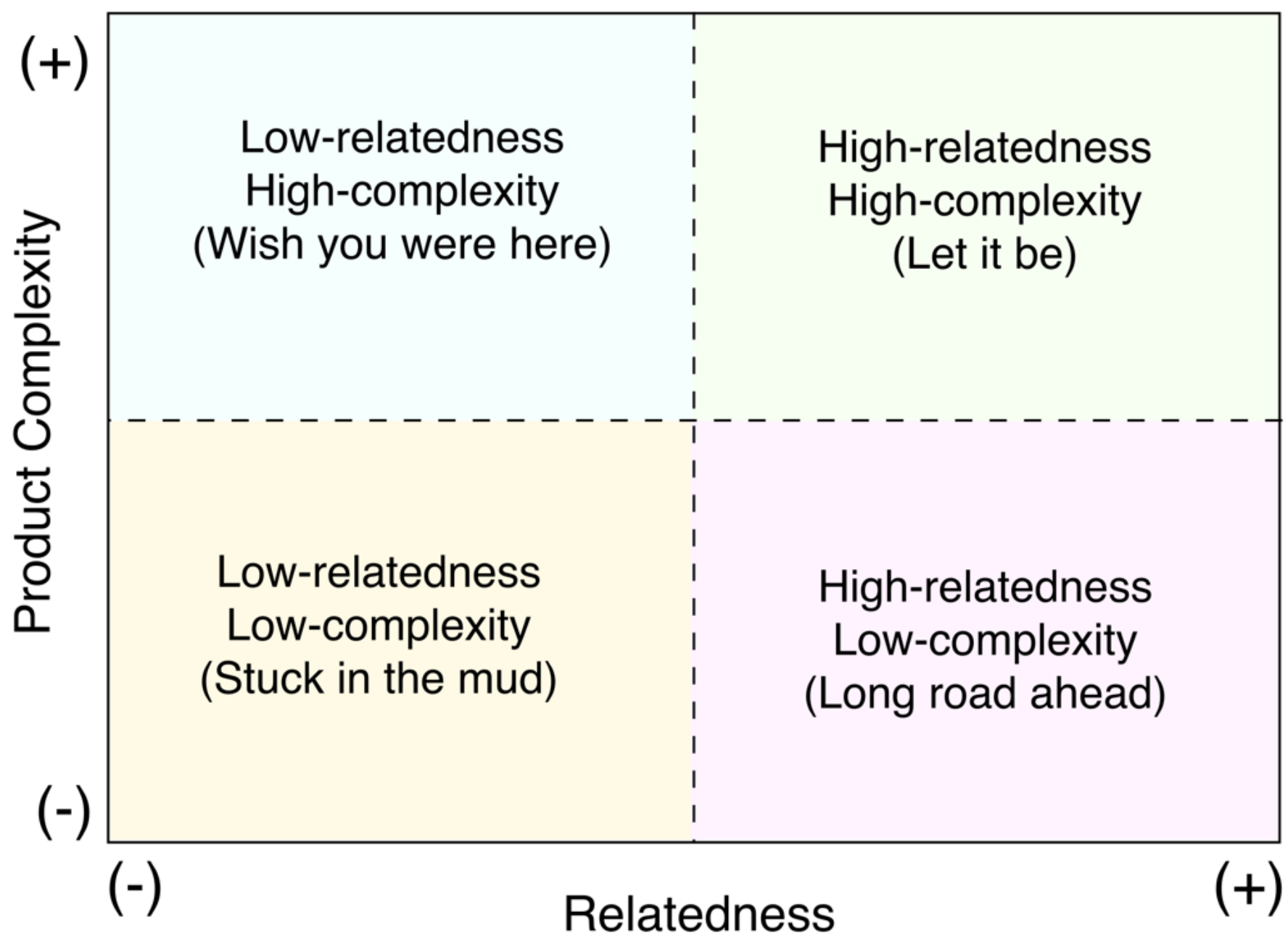

**Figure 1. Illustrative example of a relatedness-complexity diagram.**

The relatedness axis provides a measure of affinity between a location and an activity [7,10–12,33,36–38]. This is an estimate of how compatible the activity is with an economy's current productive structure (as expressed by the presence of other activities) or how "easy" it is for that economy to enter that activity (e.g. specialize its exports or employment in pharmaceuticals).*

* Technically this is a simplification, since a proper estimate of entry or exit probabilities requires defining a full model including more than relatedness type variables. For instance, when estimating changes in bilateral trade patterns at the product level, Jun et al. used a model with a total of fifteen different factors, including three types of relatedness.[37]

Complexity provides a measure of the value of each potential activity [6,39]. This is because estimates of a product's complexity are correlated with higher levels of income, and thus, serve as estimates of an activity's contribution to an economy's growth potential.

These diagrams, sometimes in combination with other criteria (e.g. transportation costs [34]), are used to identify activities in the productive frontier, which are those that are relatively high in complexity and relatedness. These activities are potentially valuable and feasible for an economy, making a relatedness-complexity diagram a simple and intuitive way to reason about strategic diversification (an idea similar to an efficient frontier in formal economic models of portfolio optimization [40–42]).

Yet, because these diagrams are based on proxy measures, and do not consider higher order effects (such as future changes in the relatedness and complexity of activities), they do not guarantee that the activities selected through them are optimal according to a more formal criterion.

Here, we address this gap by introducing an optimization method to identify diversification opportunities while considering the path dependencies implied by an economy's existing productive structure. When applied to the data, the resulting portfolios often differ from those suggested by visual inspection of relatedness–complexity diagrams and tend to place greater weight on opportunities that leverage existing capacities under explicit feasibility and cost constraints. Our approach thus adds a new target-oriented optimization layer to complement and advance the use of economic complexity methods in strategic diversification.

## The Use of Economic Complexity in Strategic Diversification

Economic complexity techniques have become common in strategic diversification efforts because of their ability to explain structural transformation processes and quantify the potential value of an economy's pattern of specialization [6,31,39,43]. Today, economic complexity ideas are part of the official industrial strategy of diversified middle-income economies, like Malaysia [44] , rich natural resource dependent economies, such as Saudi Arabia [45], and small economies, like Armenia [46]. These methods have also been applied to guide foreign direct investment efforts in Mexico [47] and have been used to justify the creation of national economic data observatories in Brazil, Mexico, and Spain among other places [48,49]. In Europe, economic complexity tools are commonly used in reports focused on regional innovation and smart diversification strategies [33,50,51] and were used to compare the relative position of Europe, the United States, and China in key technologies in an influential report prepared by a team led by Mario Draghi [52].

Most of these efforts bring economic complexity ideas into practice by using *relatedness-complexity* or *diversification frontier* diagrams (Figure 1). These diagrams combine a measure of affinity, such as relatedness, with a measure of value, such as complexity, to identify activities that are both, compatible with an economy's current patterns of specialization (high relatedness) and attractive based on their ability to generate sustainable growth (high complexity).

The use of these diagrams, however, carries important limitations.

First, relatedness is an incomplete measure of economic potential. In fact, an economy's probability of entering or exiting an activity is better approximated by models that include relatedness as one of many variables. For instance, a model of economic potential that includes also terms for an economy's specialization in each activity, trends in that specialization, and/or multiple forms of relatedness (e.g. relative relatedness [53], bilateral relatedness [37], or industry- and occupation-specific types of relatedness [10,54]). In practice, these models are more accurate than models using relatedness alone, and thus, can be said to provide a better estimate of economic potential.* But while scholars have created many models combining measures of relatedness with other predictors [10,37,55–62], the applied literature still often uses relatedness in isolation [32,33,51], risking confusing a component of an estimate of potential (relatedness) with the estimate of potential (the full model).

Second, while complexity and relatedness change over time, it is common for applied work to only use current values. For instance, by drawing diagrams using static values for the complexity of an

* Relatedness still provides valuable information in these models, since it captures the importance of spillovers among activities that are hard to capture with other variables (for instance, changes in an economy's exports of cars that can be attributed to that economy's specialization in engines, trucks, wheels, etc.).

activity instead of taking into consideration how that complexity might change over time. In some cases, changes can be substantial. Between 1962 and 2016 the product complexity of fully assembled cars (SITC 7810)—according to the product complexity index (PCI)—fell from being 17th in the ranking of all products to 208th (out of 817 activities). This is not because fully assembled cars became less sophisticated, but because many middle- and lower-income economies, such as Thailand, Morocco, and Turkey, became specialized exporters of them. In this car assembly example, a middle- or lower-income country looking to enter the car industry in the late 1980s would overestimate the increase in complexity associated with entering that sector while underestimating the competition of other new entrants.

Third, theoretical work on economic complexity [63,64] has shown that optimal diversification strategies require balancing a portfolio of related and unrelated diversification opportunities. The balance of this portfolio is expected to change with an economy's level of complexity in the form of an inverted U-shape, meaning that unrelated diversification attempts are more beneficial at an intermediate level of relatedness. Yet, while recent work [32,53] has acknowledged the need to create a balanced portfolio of related and unrelated activities, there is not yet a widely adopted principled way to estimate such a portfolio. The ECI optimization method presented in this paper provides a solution for the portfolio problem by identifying a combination of related and unrelated diversification targets that match the behavior expected in an optimal portfolio.

And fourth, it is important not to confuse the use of relatedness-complexity diagrams, or of an ECI optimization algorithm, with a method to select sectors. This is a more universal limitation that is important to note at the outset. Economic complexity methods are a useful tool for industrial policy, not because they represent an "ideal" way to select sectors, but because they are a powerful way to discard sectors chosen through less principled approaches. In the world of international development, it is common for economies to receive sectoral advice that is motivated by global trends and disregards an economy's local capacities (e.g. AI, Big Data, Sustainable Energy, Biotech). While such advice can be soothing for political elites, and even sound when there are available "windows of opportunity," [65] it can also lead to pie in the sky development efforts that fail to take local conditions into account. By providing a principled way to estimate an economy's probability of success in each sector, together with a measure of its potential value, economic complexity methods provide a rough map that can be used to ground overly optimistic targets (more details about the use of economic complexity methods in practice can be found in [32]).

During the last decade, several policy-oriented efforts have attempted to expand or improve the use of relatedness-complexity diagrams. Relatedness-complexity diagrams were introduced in the 2011 edition of The Atlas of Economic Complexity (republished later as [9]), kickstarting their use in country specific studies. This includes development reports focused on South-East Africa [66], Mozambique [67], Panama [35], and Rwanda[34], and work using patent and occupation data to look at opportunities for European regions [33,68]. This literature made progress by adding complementary information to that provided by the diagrams. For example, in the case of Rwanda, export diversification targets were first identified using a relatedness-complexity diagram and then filtered using transportation costs as an additional criterion [34]. In the case of Panama, important emphasis is placed on skilled migration [35]. In the case of Mozambique [67], a gravity model was included as a way to incorporate demand side considerations—an approach related to ITC's Export Potential Map [69] and work focused on bilateral export forecasting (e.g. [37,59]). In more recent work, Romero et al. [56] combine multiple dimensions into a score to estimate macro outcomes such as GDP per capita and employment. Yet, despite these advances, most applications still rank fixed candidate lists rather than explicitly optimizing toward a stated objective under constraints, and thus, only partially address the limitations described above.

Building on this work, we frame target sector identification as an explicit optimization toward a stated objective (e.g., complexity or growth), directly addressing the first three limitations. Our approach starts by defining a target level of economic complexity or economic growth and using an optimization method to identify a portfolio of new activities that minimizes a measure of the "effort" required to enter them. The estimate of effort is calculated as the required increase in comparative advantage needed to enter an activity that must be realized at an intermediate point in time or "steppingstone." The optimization method also includes a forward-looking model that we use to derive a future Product Complexity Index (PCI) for products, incorporating the second line of criticism.

Armed with this model, we then explore the properties of this optimization procedure using international trade data by country, and employment by industry data for cities in the United States and international patent applications data. We find that the targets identified by ECI optimization tend to leverage an economy's existing comparative advantages more than relatedness-complexity approaches, while adjusting the level of relatedness to an economy's stage of development. It also identifies a more balanced portfolio of related and unrelated activities than those identified using relatedness–complexity diagrams. In sum, the ECI optimization method adds a target-oriented optimization layer to the complexity toolkit, providing a quantitatively rigorous complement for strategic diversification.

## Methods

### The ECI Optimization Algorithm

The *ECI optimization* process begins by defining a target level of economic complexity which we operationalize by either selecting directly a value for the economic complexity index (ECI) or by choosing that value indirectly by inverting the empirically observed relationship between growth and ECI. That is, for an economy with an economic complexity of $ECI_c(t)$ at time *t*, we define a target $ECI_c^*(t + \Delta t)$. Since ECI is defined [6] as the average complexity (PCI) of the activities an economy is specialized in [6],[*] we are looking for a solution to:

$$ECI_c^* = \frac{1}{M_c^*} \sum_p M_{cp}^* PCI_p$$

Where $ECI_c^*$ is our target level of economic complexity, $M_{cp}^*$ is the future specialization matrix that we are looking to optimize (our unknown), $M_c^* = \sum_p M_{cp}^*$ is the number of activities in which economy *c* specializes in after the optimization has taken place, and $PCI_p$ are the predicted product complexity indexes of the activities at time $(t + \Delta t)$. For the standard definitions of ECI and PCI see the following review paper [39].

[*] In most applications ECI is the standardized (mean-centered, scaled) average of PCI across a country's specialized products. This normalization is a monotone affine transformation, so rankings and optimization choices are invariant to using the standardized vs. raw average. For estimation convenience, we use the unstandardized average in the intermediate steps; standardized values are reported when cross-country comparability is required.

The key mathematical question here is how to determine the specialization matrix $M^*_{cp}$. That requires defining an optimization criterion or cost function. Here we look for the $M^*_{cp}$ that achieves the target level of economic complexity ($ECI^*$) while minimizing the sum of the increase in comparative advantages at an intermediate timepoint, or "steppingstone." In principle, we could choose other constraints, such as minimizing the added volume of exports in the case of trade or minimizing the increase in added payroll for industries. For illustration and clarity purposes, we minimize the sum of comparative advantages as our criterion, since it provides reasonable results and avoids some limitations of other optimization constraints. For instance, minimizing the added volume of employment or exports biases the optimization process towards smaller activities.

As usual, we define $M^*_{cp}$ as a binarized measure of a specialization matrix $R^*_{cp}$, which is an estimate of revealed comparative advantage* (RCA),

$$R_{cp} = \frac{X_{cp} X}{X_c X_p}$$

where $X_{cp}$ is a measure of output, volume, or value added (depending on availability, e.g. total exports of country $c$ in product $p$, total payroll paid by city $c$ in industry $p$, etc.). Also, we use Einstein's notation where muted indexes were added over (e.g. $X_c = \sum_p X_{cp}$).

That is, $M^*_{cp} = 1$ if $R^*_{cp} \geq 1$ and 0 otherwise. We then calibrate the coefficients of our forecast model by using a linear regression of the form:

$$\mathrm{r}_{cp}(t + \Delta t) = b_1 \mathrm{r}_{cp}(t + \tau) + \; b_2 \mathrm{r}_{cp}(t) + b_3 \omega_{cp}(\mathrm{t}) + \; b_4 \widetilde{\omega}_{cp}(\mathrm{t}) + b_0 + e_{cp}(t) \qquad (2)$$

where

---

* where $X_{cp}$ is an output matrix, summarizing the output of location $c$ in activity $p$, and $X_c = \sum_p X_{cp}$, $X_p = \sum_c X_{cp}$, and $X = \sum_{c,p} X_{cp}$.

$$\mathrm{r}_{cp}(t) = \log(R_{cp}(t) + 1)$$

and $R_{cp}(t)$ is the specialization (RCA), $t$ is the initial time point, $t + \Delta t$ is the final time point, and $\tau < \Delta t$ is the steppingstone time point. Also, $\omega_{cp}(\mathrm{t})$ is the relatedness of location $c$ in activity $p$, $\widetilde{\omega}_{cp}(\mathrm{t})$ is the relative relatedness of the same location-activity pair[53], $e_{cp}(t)$ is the error term, and $b_0$ is the intercept of the model.

We calibrate the coefficients in equation (2) using historical data, by assuming that coefficients are different for entry models (defined as $R_{cp}(t) < 1$ ) and exit models (defined as $R_{cp}(t) \geq 1$ ). For instance, for a model starting on the year $t$ =2012, with a five-year steppingstone ($\tau$=5) and a ten-year horizon ($\Delta t$=10), we calibrate the coefficients by regressing the output in 2022 with a steppingstone in 2017. We justify this calibration by showing that the coefficients determined through this process are similar regardless of the starting year. In fact, as we should expect, the coefficients depend more on the duration of the steppingstone than on the final year. For instance, a 10-year forecast with a 5-year steppingstone has similar coefficients when choosing 2002 or 2012 as the starting year but has different coefficients than a 10-year forecast with a steppingstone on year 2.

In Table 1, we show two examples of regression analyses predicting output in 2022, using a 5-year steppingstone in the upper panel and a 10-year steppingstone in the lower panel. In both cases, we find that the models including all four explanatory variables (column 3 for entry models and column 6 for exit models) have the highest explanatory power. Additionally, we observe that all explanatory variables show a positive and statistically significant relationship with the output variable.

**Table 1. Entry and Exit Model for Target Year 2022**

| | *Dependent variable: log(1 + RCA) (2022) [in 5 years]* | | | | | |
|---|---|---|---|---|---|---|
| | | Entry | | | Exit | |
| | (1) | (2) | (3) | (4) | (5) | (6) |
| log of RCA (2017) | 0.648*** | | 0.636*** | 0.889*** | | 0.874*** |
| | (0.003) | | (0.003) | (0.005) | | (0.005) |
| log of RCA (2012) | 0.283*** | | 0.215*** | 0.007 | | 0.022*** |
| | (0.004) | | (0.005) | (0.006) | | (0.006) |
| Relatedness (2012) | | 0.878*** | 0.180*** | | -0.653*** | 0.284*** |
| | | (0.007) | (0.007) | | (0.042) | (0.022) |
| Relative Relatedness (2012) | | 0.043*** | 0.014*** | | 0.271*** | 0.023*** |
| | | (0.001) | (0.001) | | (0.006) | (0.003) |
| Observations | 134948 | 134948 | 134948 | 24217 | 24217 | 24217 |
| $R^2$ | 0.459 | 0.140 | 0.465 | 0.754 | 0.087 | 0.757 |
| Adjusted $R^2$ | 0.459 | 0.140 | 0.465 | 0.754 | 0.087 | 0.757 |

| | *Dependent variable: log(1 + RCA) (2022) [in 10 years]* | | | | | |
|---|---|---|---|---|---|---|
| | | Entry | | | Exit | |
| | (1) | (2) | (3) | (4) | (5) | (6) |
| log of RCA (2012) | 0.641*** | | 0.627*** | 0.823*** | | 0.817*** |
| | (0.003) | | (0.003) | (0.006) | | (0.006) |
| log of RCA (2002) | 0.284*** | | 0.196*** | 0.025*** | | 0.020*** |
| | (0.005) | | (0.006) | (0.006) | | (0.007) |
| Relatedness (2002) | | 1.056*** | 0.249*** | | -0.506*** | 0.044 |
| | | (0.010) | (0.009) | | (0.048) | (0.031) |
| Relative Relatedness (2002) | | 0.038*** | 0.014*** | | 0.227*** | 0.031*** |
| | | (0.001) | (0.001) | | (0.006) | (0.004) |
| Observations | 107669 | 107669 | 107669 | 21121 | 21121 | 21121 |
| $R^2$ | 0.417 | 0.126 | 0.423 | 0.660 | 0.062 | 0.661 |
| Adjusted $R^2$ | 0.417 | 0.126 | 0.423 | 0.660 | 0.062 | 0.661 |

Note: *p<0.1; **p<0.05; ***p<0.01

Since there are multiple choices for a starting year, given a steppingstone and a horizon, we average the model's estimates across all possible initial years $t$, yielding a more reliable parameter set (see the Supplementary Material). Using these coefficients, we set up a second equation where we allow the comparative advantages in the steppingstone year to change from those in the initial year by an amount equal to an unknown matrix $W_{cp}$. That is, we set up the model:

$$r_{cp}(t+\Delta t) = b_1 \log(R_{cp}(t) + W_{cp} + 1) + b_2\, \mathrm{r}_{cp}(t) + b_3 \omega_{cp}(\mathrm{t}) + b_4 \widetilde{\omega}_{cp}(\mathrm{t}) + b_0 + e_{cp}(t) \quad (3)$$

which we can solve algebraically for $W_{cp}$ when setting $R_{cp}(t+\Delta t) = 1$ (or $r_{cp}(t+\Delta t) = \log(2)$ for the target year. That is, we estimate the $W_{cp}$ that correspond to an entry, making $W_{cp}$ an estimate of the change in specialization (the added $RCA$) that an economy needs to achieve by the steppingstone year (t+ $\tau$) for the model to predict that this economy will specialize in an activity by year $t+\Delta t$.

Finally, we use the coefficients in $W_{cp}$ in a binary optimization program (a.k.a. 0-1 integer optimization)[70,71] . The program minimizes the sum of the weights in $W_{cp}$ for new specializations ($M_{cp}(t) = 0 \rightarrow M_{cp}(t+\Delta t) = 1$) that do not require a boost from $W_{cp}$, subject to reaching an economic complexity equal or larger than the target. That is, we look for:

$$\frac{\sum_p M_{cp}(t+\Delta t) PCI_p}{\sum_p M_{cp}(t+\Delta t)} \geq ECI_c^*$$

while minimizing:

$$\min(\sum_p W_{cp} M_{cp}^*(t+\Delta t))$$

where $PCI_p$ is obtained from the specialization matrix predicted by equation (3) when setting $W_{cp} = 0$ for time $t+\tau$.* We justify this assumption by noting that while the *PCI* values would change after the optimization, here we optimize only one economy, which should translate into a

* Note that we can change the ECI constraint in our optimization to any other target related to specialization (e.g., diversity, Fitness, etc.) as long as it is a linear function of the specialization patterns.

minor impact on $PCI_p$. Thus, we approximate the complexity of activities by using the future $PCI$ implied by the model's forecast.

In sum, $M^*_{cp}$ provides the portfolio of activities that optimally reaches a target level of economic complexity while minimizing the increase in comparative advantage needed to achieve that target at the intermediate time point.

To avoid implausible diversification paths, we set a data-driven complexity ceiling conditional on current ECI. At the baseline, we group countries into ECI bands of width 0.2 (from −2.5 to 2.5); for each band we pool the PCI of all products currently exported with RCA ≥ 1 by countries in that band and take the boxplot upper whisker as the band's ceiling (Q3 + 1×IQR). We then smooth these ceilings with an OLS fit on band midpoints to obtain a continuous ECI→PCI ceiling; a country may enter only products with PCI at or below this ceiling, which we compute once at baseline and hold fixed during optimization (allowing some capability stretch while ruling out "giant leaps").

Notice that our approach assumes that one economy optimizes at a time (no general equilibrium effects with multiple economies optimizing simultaneously while knowing that other economies are also attempting to optimize). This simplifies the problem but does not fully capture general equilibrium effects. Despite this limitation, ECI optimization should enhance the application of economic complexity methods by providing a more strategic and forward-looking approach to diversification.

Figure 2 summarizes the ECI optimization process and its benefits over standard complexity-relatedness diagrams. First, instead of relying solely on relatedness, it incorporates a more detailed and generalizable measure of the effort needed to add an activity, represented by the additional $RCA$ required in the steppingstone year in the context of a multivariate model. Second, ECI optimization adopts a forward-looking perspective by integrating forecasts of changes in both RCA and complexity values over time. This dynamic view helps economies avoid overestimating potential complexity gains from industries that may become more ubiquitous as other countries also develop specialization in them.

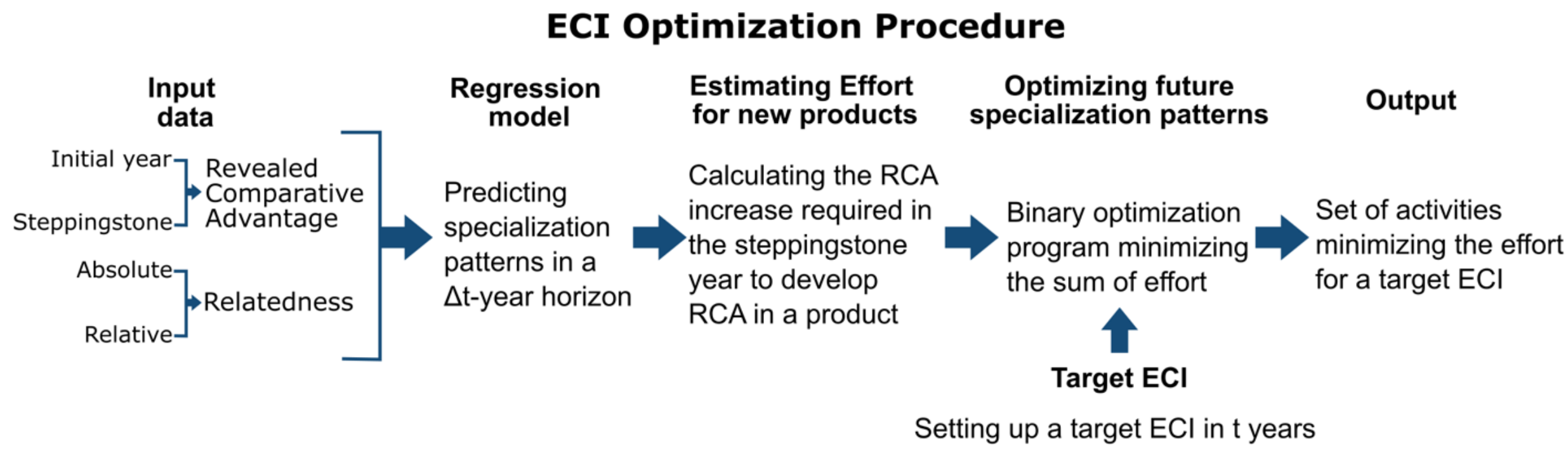

**Figure 2. ECI Optimization Systems Diagram.**

## Data

We use three datasets to evaluate the performance of ECI Optimization. The main text presents results with international trade data; the Supplementary Materials (SM) report robustness using (i) subnational U.S. payroll data and (ii) patent applications data.[*]

In the main manuscript we present results using international trade data, disaggregated by the HS4 Classification (Rev. 1992), spanning 1998–2022 and sourced from the Observatory of Economic Complexity (oec.world). To reduce noise from year-to-year fluctuations, we apply a four-year moving average to exports (e.g., "2022" averages 2019–2022). For each year, we include only countries with exports > USD 1 billion and population > 1 million, and we exclude products with annual trade < USD 500,000 to minimize noise from small flows.

In the SM, we report results using subnational U.S. payroll data by metropolitan area from 2003–2022, taken from the United States Census Bureau's County Business Patterns (https://www.census.gov/programs-surveys/cbp.html) and from WIPO's Patent Cooperation

[*] The data and code that can be used to reproduce the results of the paper are available at: https://doi.org/10.7910/DVN/RXDCI1.

Treaty (PCT) system. For US payroll data, we compute four-year averages of payrolls to smooth volatility and retain only metropolitan areas with total payroll > USD 100,000 and activities with total payroll > USD 150,000, which reduces spurious variation coming from small units of observation. For WIPOs PCT data, patents are attributed to the inventors' country of residence and classified into CPC 4-digit technology classes; when inventors span multiple countries, each inventor residence is credited once. We use data from 1999 to 2021 to compute four-year moving averages of applications by country–class and, for each year, retain only classes with 5+ applications.

**Regression model parameter estimation**

We use an Ordinary Least Squares (OLS) approach to estimate the parameters defined in equation (2), We estimate the model separately for each initial year within a fixed timeframe $\Delta t$ and steppingstone $\tau$. To improve stability and reduce noise from year-to-year fluctuations, we average the OLS estimates across all possible initial years $t$ (for a fixed steppingstone $\tau$), yielding a more reliable parameter set. This approach accommodates time-heterogeneous relationships, keeps estimation tractable given the scale of the country–product panel, and is compatible with the closed-form computation of effort $W_{cp}$ used in the optimization.*

Our estimation differentiates between entry and exit models. Entry models use data only on activities where $R_{cp}(t) < 1$, i.e., modeling only the activities in which locations could potentially establish new specializations at time $t + \Delta t$. In contrast, exit models consider only data on activities where $R_{cp}(t) \geq 1$, capturing only activities in which a location can lose specialization.

In the empirical analysis presented throughout the manuscript we set the timeframe $\Delta t$ to 10 years, and the steppingstone $\tau$ to 5 years. Also, as a starting point for the future predictions we use the

* In the Supplementary Material we show that the estimated parameters are quite stable over time (see Fig. S5, Fig. S10, and S15). Under such stability, a pooled fixed-effects panel with country–product and year effects would yield a slope numerically very close to the simple average of our year-specific OLS slopes. We adopt the year-by-year estimation primarily for computational practicality: it is straightforward to run on standard hardware given the size of the country×product panel, without specialized high-dimensional fixed-effects solvers. In practice, both approaches should deliver substantively similar parameters as long as the parameters are stable over time.

latest year, 2022, meaning that we use our models to predict the geography of economic activities in 2023, and use ECI Optimization to select optimal portfolios for the 2027 steppingstone.

## Results

### Example

We begin by illustrating the ECI Optimization method by drawing an effort ($W$)-complexity ($PCI$) diagram (Figure 3) for Vietnam. In this diagram we plot the expected Product Complexity Index ($PCI_p$) (in the target year) for each potential new activity $p$, alongside the effort $W_{cp}$ required at the steppingstone year for an economy to reach a "positive" level of specialization ($R_{cp}(t + \Delta t) \geq 1$).

The optimization process begins by selecting activities in the upper-left quadrant (Figure 3 a), which represent the most efficient choices in terms of the tradeoff between complexity and effort. But how many activities do we need to choose to reach an ECI target? And how do we trade-off between low effort and high complexity?

As we raise the target ECI (Figures 3b and 3c), we must gradually include more activities—some with higher required effort (moving rightward) and some with slightly lower complexity (moving downward). This sequential inclusion reflects the trade-off between increasing complexity and managing the additional effort required.

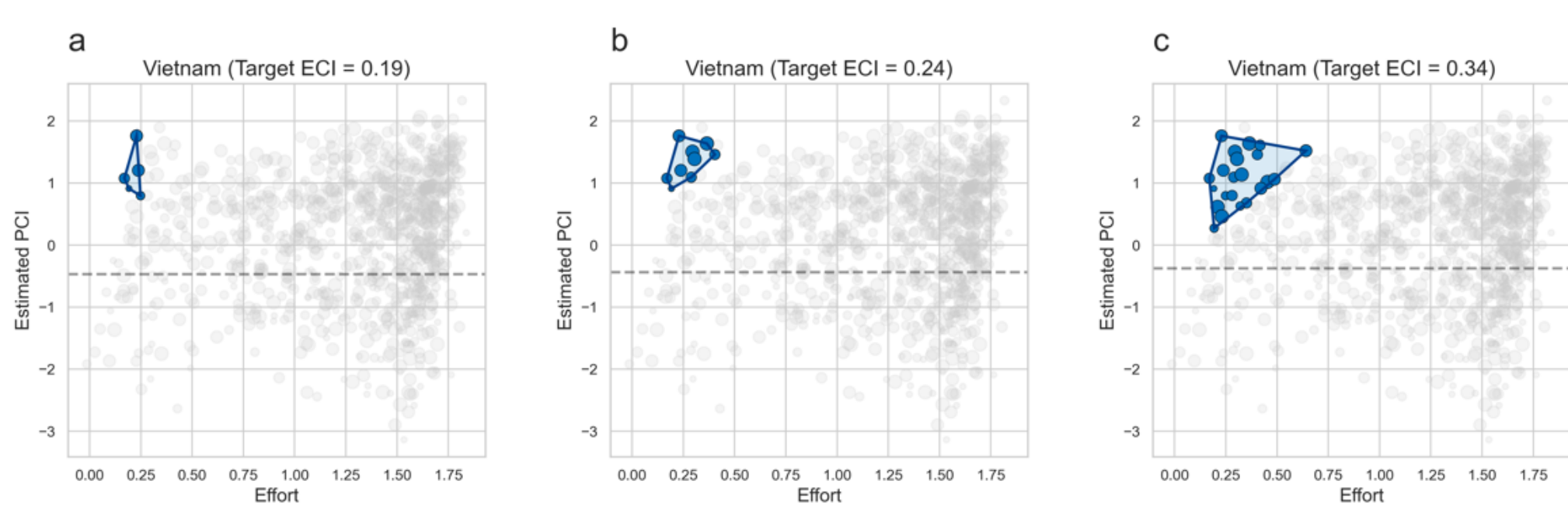

**Figure 3. The ECI Optimization Method. a)** ECI Optimization effort-complexity diagram for Vietnam's exports to reach a target ECI of 0.19 (starting from an ECI of 0.14). The selected activities to reach a target ECI are highlighted in blue. **b)** Same as **a)**, only for a target ECI of 0.24. **c)** Same as **a)**, only for a target ECI of 0.29.

## Properties of ECI Optimization

We explore the properties of the ECI optimization method by comparing it with a benchmark that approximates a common practice in the use of relatedness-complexity diagrams, that is, identifying activities that are relatively high in complexity and relatedness. We implement this benchmark by first min-max normalizing the relatedness and complexity (PCI) of each activity (to a 0 to 1 scale) and then taking the product of these values. Then, we select the *N* activities with the highest score needed to reach the target ECI.

For this example, we use international trade data disaggregated by the HS4 classification (Revision 1992), downloaded from the Observatory of Economic Complexity (oec.world) and focus on using 2022 data to model the exports in 2032. That means setting 2027 as a steppingstone year. For each country, we target an increase in ECI of 0.1 standard deviations ($ECI(t + \Delta t) = ECI(t) + 0.1$)). More details about the data are available in the appendix.

Figure 4 compares ECI optimization with the *relatedness-complexity* diagram benchmark.

Figure 4a looks at the average Revealed Comparative Advantage (RCA) in 2022 for the products suggested by both methods. This shows that *ECI optimization* suggests products where countries currently have higher specialization compared to those suggested by the benchmark. This is a reasonable outcome, since ECI optimization minimizes the additional RCA required to achieve a target ECI. But this is also an important difference between the two methods, as it shows that ECI optimization focuses on activities that leverage existing capacities (as expressed in products close to full levels of specialization (e.g. RCA~0.8)).

Figure 4b compares the two methods by looking at their average relative relatedness (relatedness centered on a country's mean). Compared to the benchmark, ECI optimization selects products that are relatively more unrelated for lower and middle complexity economies. The relationship defined by the ECI optimization choices also has a curvature resembling the one observed for the optimal diversification model introduced by Alshamsi et al.,[63,64] which shows that effective diversification strategies require targeting both related and unrelated activities. This is confirmed by the inset of Figure 4b, which shows that the standard deviation of the recommended activities drastically increases for countries with a medium-high value of ECI (ECI~0.5). Hence, ECI optimization may endogenously lead to a more balanced portfolio of related/unrelated activities in accordance with models of optimal portfolio diversification, balancing the pursuit of complex products with the feasibility associated with their relatedness.

Figure 4c plots the number of suggested products as a function of an economy's diversity (the number of activities in which a country had RCA>1 in 2022). Here, both methods behave similarly, showing that, while they recommend different activities, they recommend a similar number of them. Still, ECI optimization suggests a slightly lower number of new activities compared to the benchmark model.

Finally, Figure 4d plots the minimum value of additional export volume required to reach comparative advantage in the suggested products as a function of diversity in 2022. This additional export volume can be considered an alternative measure of the effort or cost needed to develop new specializations. We find that the ECI optimization method suggests slightly lower additional volume compared to the benchmark model (around USD 1B on average), suggesting that it effectively balances the trade-off between the feasibility of expanding export volumes and the potential gains in economic complexity.

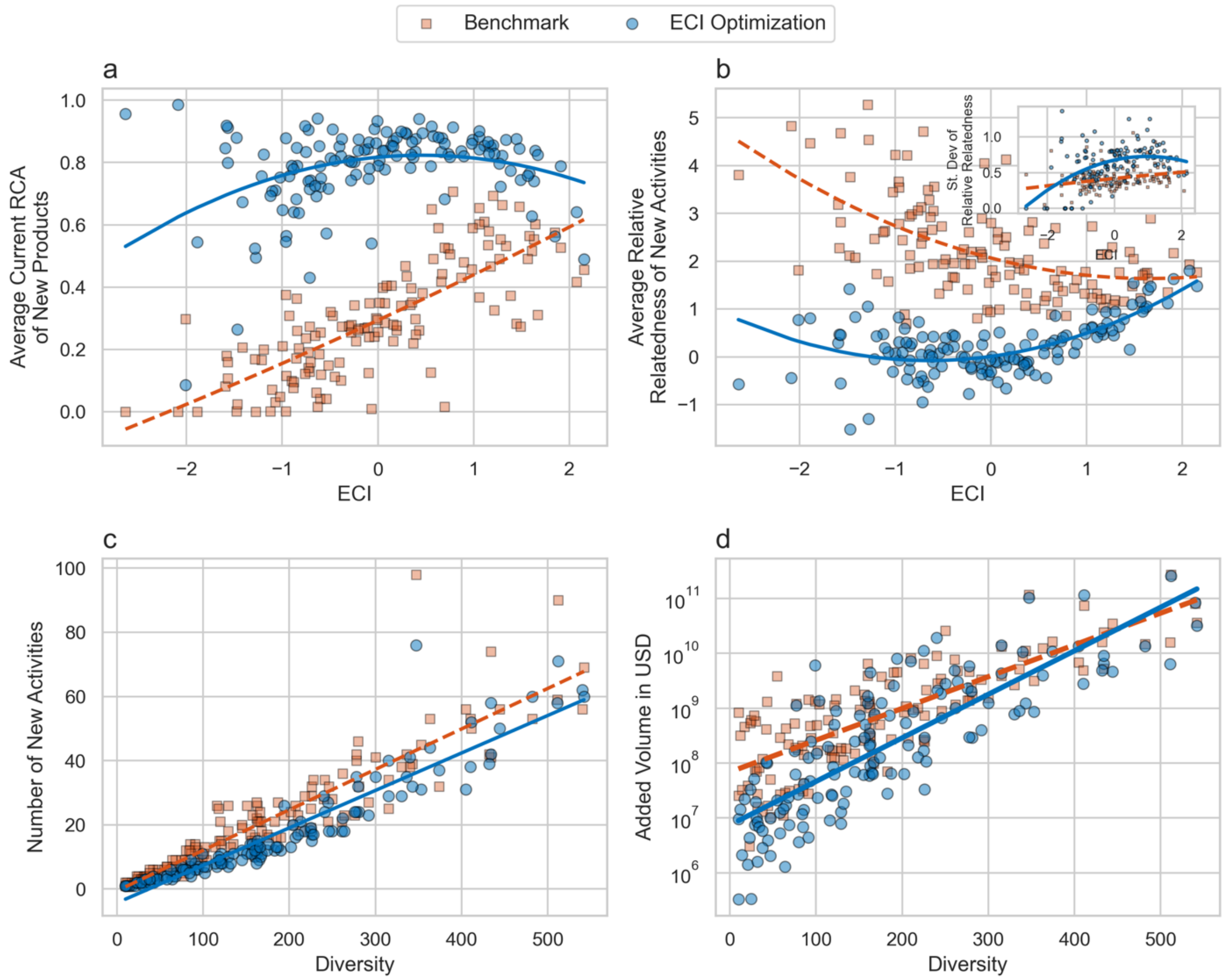

**Figure 4. Properties of ECI Optimization in International Trade Data. a)** Average RCA (in 2022) of the products suggested by ECI optimization and the benchmark model as a function of a country's ECI in 2022. **b)** Average relative relatedness of the suggested products as a function of a country's ECI in 2022. The inset plot shows the variance of the relative relatedness of the suggested products. **c)** The number of suggested activities as a function of the initial diversity (number of activities in which the country had comparative advantage in 2022). **d)** Estimated added export volume to gain comparative advantage in the suggested products as a function of the initial diversity. **a-d)** For each country we assume an increase of 0.1 of the ECI value in 2022. All scatter charts include a quadratic fit line.

In the supplementary material, we repeat this analysis using the USA's MSA payroll by industry data and international data on patent applications from WIPO's PCT system, finding similar results. Hence, our results are robust to the granularity of the data (national vs. subnational) and the classification of activities (product exports vs. industry payroll vs. patents application). Overall, these results suggest that ECI optimization recommends activities where economies already have some comparative advantage while balancing relatedness with the required level of added output.

### Targeting Economic Growth

Finally, we use ECI optimization to target a level of economic growth.* Differently from scorecard approaches[56], that select products first and then project the growth they imply, we invert the sequence: we start from a target growth rate, back out the ECI consistent with it, and then identify the minimal-effort portfolio to reach that ECI.

We start by constructing the following panel growth regression using ECI as an explanatory variable:

$$\log \frac{GDPpc_{t+10}}{GDPpc_t} = a_1 ECI_{ct} + a_2 \log GDPpc_t + a_3 ECI_{ct} \times \log GDPpc_t + \gamma_t + u_{ct}. \tag{4}$$

Here, the dependent variable is the annualized 10-year growth rate of GDP per capita (in PPP constant 2021 USD) of a country (we consider two periods 1999–2009 and 2009–2019).† Besides ECI, we use two additional explanatory variables. First, we use a z-score normalized value for the log of initial GDP per capita (normalized across our sample of countries for each year), capturing Solow's idea of economic convergence.[72]‡ Second, we use the interaction of ECI with the z-score of the initial log of GDP per capita, capturing the idea that the contribution of economic complexity to future economic growth is stronger for lower income economies.[9] We also use period fixed effects to account for omitted variables that vary over the two decades and may impact economic growth.§

* Note that the model can be adapted to any other policy target (e.g., inequality, sustainability) by changing the dependent variable in equation (4).

† We opt to model growth in terms of annualized changes of GDP per capita PPP (constant 2021 USD) because this approach captures cross-country variations in income levels and cost structures, providing a more comparable measure of growth. These predictions can be translated into standard GDP growth rates by incorporating population growth, or into simpler GDP per capita growth rates by discounting the adjustments for PPP.

‡ This z-score transformation helps us account for the non-stationary nature of GDP per capita and provide a consistent prediction.

§ We do not use country fixed effects since our panel has only two time periods. In this case, country fixed effects would take almost all of the degrees of freedom.

Once we estimate the parameters of the model (see Table S1 in the Supplementary Information) we can set a target growth rate and invert the equation to find the ECI which is compatible with that level of growth. Here, we use two countries with a similar ECI in 2022 as examples: Thailand (ECI = 0.98) and Mexico (ECI = 0.99).

We start by using our steppingstone model to estimate the export structure of these countries and their corresponding ECI in 2032 without any optimization. Specifically, we (i) forecast product complexity to 2032 using the PCI projection in eq. (2); (ii) project specialization using the stepping-stone model in eq. (3) with $W_{cp} = 0$ (no policy shock), yielding $R_{cp}(t + \Delta t)$ and $M_{cp}(t + \Delta t) = 1[R_{cp} \geq 1]$; and (iii) compute 2032 ECI from the projected export matrix. Our model estimates that in 2032 Thailand will have an ECI = 0.933, whereas Mexico will have ECI = 0.926, meaning that the countries may experience a slight decrease in their *ECI*, remaining largely at a similar level of complexity.

Next, we use our growth regression model to provide an estimate for the expected economic growth rate that supports the predicted ECI. Even though the countries have a similar ECI, they do have a slightly different expected growth rate (assuming that the other economic conditions, will remain the same as in the last time period, e.g. fixed effects and relative GDP per capita): Thailand has an expected annualized growth rate of 3.23%, whereas Mexico has 3.15%. This is because they have a different starting level of GDP per capita.

Then, we ask the question of which is the optimal portfolio according to ECI optimization for these countries to increase their growth potential to 3.5%? To answer this, we invert the growth equation (3) and recover an expected value of ECI that supports the targeted growth. For Thailand, we find that ECI = 1.225, whereas for Mexico ECI = 1.287.

We use these values as target ECI's in ECI optimization and suggest new activities for the countries. Figures 5 a and b depict, respectively, the results for Thailand and Mexico using effort-complexity diagrams. In the figures we also highlight the results from the benchmark model. We observe that ECI optimization yields significantly different predictions compared to those of the benchmark model. In each case, ECI optimization suggests the products located in the top left of the diagram, which are “high gain” and “low effort.” By contrast, the benchmark model suggests many more products are located on the top right, meaning that they require high effort.

In Figure 5c, we summarize the ECI Optimization output for Thailand in a table, listing the products in the order of the first target growth rate they recommended, up to the target of 3.5%. For lower targets, the ECI Optimization method suggests products like 8703: Cars and 7320: Iron Springs, which Thailand can develop with relatively low effort. As the target ECI rises to a value that corresponds to an annualized growth rate of 3.5%, the list expands to include higher-effort, higher-complexity products such as 9031: Other Measuring Instruments.

By comparison, the suggested products for Mexico (shown in Figure 5d) differ significantly from those of Thailand. For lower ECI targets, the ECI Optimization method recommends goods like 8547: Metal Insulating Fittings and 8485: Additive Manufacturing Machines, while higher ECI targets introduce additional products such as 8427: Forklifts. This contrast highlights how the method adapts recommendations to fit the economic contexts of different economies, accounting for each economy’s predicted future structure and existing comparative advantages.

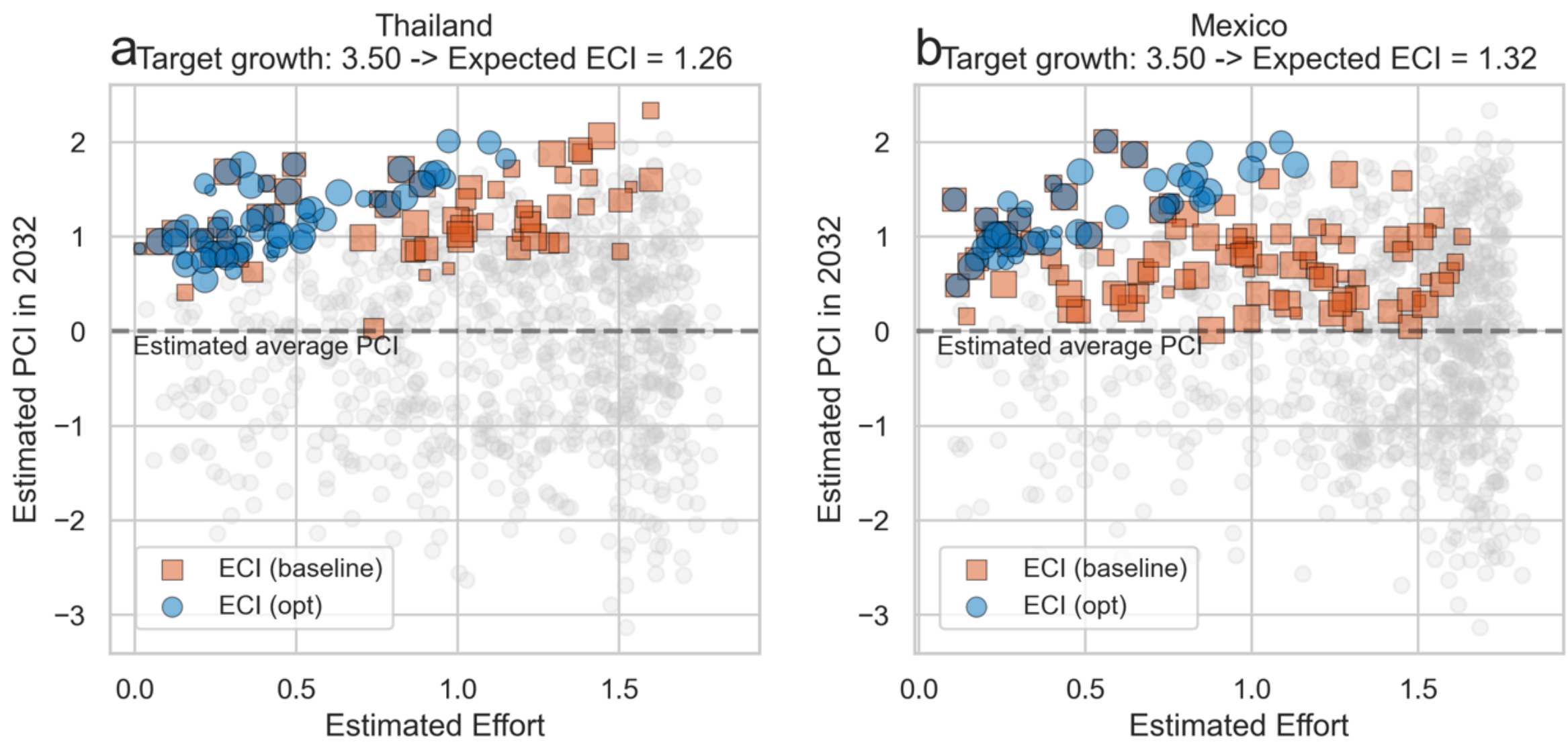

c

Thailand

| Product | $R_{cp}$ | $\omega_{cp}$ | $\tilde{\omega}_{cp}$ | PCI | ECI | Growth(%) |
|---|---|---|---|---|---|---|
| 7315 | 0.940 | 0.245 | -0.829 | 1.561 | 0.98 | 3.241 |
| 8703 | 0.964 | 0.265 | 0.402 | 0.943 | 0.98 | 3.241 |
| 8532 | 0.859 | 0.315 | 3.492 | 1.034 | 0.98 | 3.241 |
| 7320 | 0.935 | 0.266 | 0.465 | 1.109 | 0.98 | 3.241 |
| 7105 | 0.900 | 0.257 | -0.090 | 1.491 | 0.98 | 3.241 |
| ... | ... | ... | ... | ... | ... | ... |
| 3906 | 0.771 | 0.265 | 0.409 | 1.203 | 1.13 | 3.381 |
| 8506 | 0.773 | 0.289 | 1.863 | 0.801 | 1.13 | 3.381 |
| 8455 | 0.696 | 0.253 | -0.317 | 1.294 | 1.13 | 3.381 |
| 8428 | 0.675 | 0.257 | -0.071 | 1.284 | 1.13 | 3.381 |
| 8535 | 0.835 | 0.270 | 0.716 | 0.769 | 1.13 | 3.381 |
| ... | ... | ... | ... | ... | ... | ... |
| 2813 | 0.785 | 0.246 | -0.782 | 0.794 | 1.26 | 3.503 |
| 8461 | 0.302 | 0.244 | -0.892 | 1.823 | 1.26 | 3.503 |
| 2926 | 0.714 | 0.249 | -0.590 | 0.946 | 1.26 | 3.503 |
| 8483 | 0.477 | 0.257 | -0.089 | 1.419 | 1.26 | 3.503 |
| 8208 | 0.426 | 0.242 | -1.028 | 1.611 | 1.26 | 3.503 |

d

Mexico

| Product | $R_{cp}$ | $\omega_{cp}$ | $\tilde{\omega}_{cp}$ | PCI | ECI | Growth(%) |
|---|---|---|---|---|---|---|
| 8547 | 0.912 | 0.199 | 1.143 | 1.183 | 0.98 | 3.189 |
| 8485 | 0.980 | 0.201 | 1.228 | 1.392 | 0.98 | 3.189 |
| 8484 | 0.905 | 0.178 | -0.005 | 1.373 | 0.98 | 3.189 |
| 6810 | 0.973 | 0.201 | 1.236 | 0.482 | 0.98 | 3.189 |
| 6815 | 0.905 | 0.200 | 1.199 | 0.955 | 0.98 | 3.189 |
| ... | ... | ... | ... | ... | ... | ... |
| 8423 | 0.891 | 0.153 | -1.411 | 0.990 | 1.13 | 3.327 |
| 8707 | 0.860 | 0.199 | 1.159 | 0.911 | 1.13 | 3.327 |
| 8483 | 0.753 | 0.193 | 0.815 | 1.419 | 1.13 | 3.327 |
| 8205 | 0.570 | 0.164 | -0.775 | 1.651 | 1.18 | 3.373 |
| 5602 | 0.895 | 0.185 | 0.396 | 0.738 | 1.18 | 3.373 |
| ... | ... | ... | ... | ... | ... | ... |
| 8480 | 0.486 | 0.176 | -0.121 | 1.475 | 1.32 | 3.502 |
| 9013 | 0.540 | 0.156 | -1.235 | 1.390 | 1.32 | 3.502 |
| 8433 | 0.691 | 0.198 | 1.063 | 1.012 | 1.32 | 3.502 |
| 4806 | 0.510 | 0.172 | -0.367 | 1.400 | 1.32 | 3.502 |
| 8207 | 0.322 | 0.174 | -0.210 | 1.759 | 1.32 | 3.502 |

**Figure 5. Applying ECI Optimization for Diversification. a)** Effort-Complexity diagram for Thailand to increase its 2032 ECI to a value that corresponds of a 3.5% annualized GDP per capita growth rate. **b)** Same as **a**, for Mexico. **c)** List of suggested products for Thailand for a sequential increase of ECI to a value that corresponds to a 3.5% annualized GDP per capita growth rate. The products are ordered by the first target ECI at which they are suggested. **d)** Same as **c**, just for Mexico. **a-d**. The optimization is done using data classified according to the HS4 Rev. 1992 classification.

## Discussion

During the last decade and a half, economic complexity applications have built gradually on the use of relatedness-complexity diagrams. Here, we complement these diagrams by proposing an optimization method that helps identify activities that are compatible with reaching a target level of economic complexity with minimal effort. We find that the activities identified through this method are more balanced in terms of mixing related and unrelated activities than those identified using relatedness-complexity diagrams, indicating that ECI optimization represents a possible solution to the portfolio optimization problem of strategic diversification.

Yet, our approach is not without limitations.

First, in this paper we operationalize the framework using ECI and growth in GDP per capita as objectives. This is an incidental choice that could be extended to other targets by changing the dependent variable in the regression model (4) (e.g., sustainability, inequality). Nonetheless, optimizing any index raises a Goodhart-law type concern: when a measure becomes a target, it can lose its diagnostic value [73–75]. By explicitly optimizing for ECI, we risk altering the patterns the index was designed to summarize, potentially weakening its usefulness as a structural indicator.

Second, ECI and growth are particularly volatile targets, especially for small and/or less-diversified economies. In these cases, ECI can fluctuate sharply because adding or dropping one specialization mechanically shifts the average PCI, which could be caused by changes in commodity prices in the case of mineral resource exports. We view this primarily as sensitivity to composition rather than measurement error; to mitigate it without changing the framework, the model could be easily adapted to target a multi-year average ECI rather than a single-year value—an approach that can be applied analogously to alternative objectives.

A third limitation concerns the general application of economic complexity as a tool for policy. While economic complexity methods are commonly used to suggest activities that could drive economic growth, it is crucial to remember that the index itself is a 'philosophically positive' measure [32,76]. That is, ECI reflects the presence of multiple capabilities or inputs [6,77], and thus, applications of ECI optimization should not stop at the generated recommendations but should also delve into the broader dynamics that shape these outcomes. For example, Economic complexity, as used in Malaysia's New Industrial Master Plan [44], is a mission acting as a rallying flag hoping to coordinate multiple activities towards a common goal. An effective application of ECI-based strategies requires acknowledging the spatial and organizational aspects of knowledge growth and diffusion that underpin economic structures [55,78–82]. This includes, for instance, fostering connections with global leaders in specific sectors or cultivating institutions that can support these new activities. Without these foundational elements, the benefits of ECI recommendations may be short-lived.

Fourth, our illustration favors parsimony: a compact diversification model (two density measures plus specialization) and a linear stepping-stone specification make the optimization transparent and computationally tractable as a 0–1 linear program [10,37,54,83–87]. The framework is flexible, but not fully model-agnostic: to keep the optimization in this tractable form, the stepping-stone equation must be invertible in $W_{cp}$ so that effort can be computed from (3) in closed form (or via a simple one-dimensional inversion). This requirement motivates the log-linear form we use, which delivers an analytic $W_{cp}$. Richer estimators (PPML, non-linear splines/interactions, hierarchical panels, or ML surrogates) are natural extensions, provided they preserve invertibility; otherwise, one can precompute $W_{cp}$ numerically for each location-activity pair. Similarly, we hold relatedness density fixed at the baseline year and some products may be attractive only within finite windows of opportunity. As densities and priorities shift with specialization, iterative or rolling-window updates would better capture path dependence. These dynamics could be incorporated into a time-staged extension that addresses when to enter, not only what to enter.

Fifth, our optimization model primarily considers a single dimension of economic activity—such as international trade data (or patents applications) or industrial activities within MSAs. Economic complexity, however, is inherently multidimensional [31]. It encompasses not only the sophistication of products or industries but also the complexity of other structures like technologies [88,89], research [50], and the digital economy [90,91]. These additional dimensions are crucial for an economy's potential for inclusive and sustainable growth. By limiting our analysis to a singular dimension, we risk overlooking these critical aspects, thereby restricting the comprehensiveness and applicability of our diversification recommendations.

Finally, a key limitation of this approach is that it does not fully account for general equilibrium effects, where multiple economies optimize simultaneously while considering how others are also optimizing. While RCA and ECI help mitigate this issue by being relative measures, implying that the global average adjusts as countries change their specializations, our model does not explicitly account for interdependencies in decision-making across countries. Addressing this would require a game-theoretic or agent-based extension of the framework, which we leave for future research.

Despite these limitations, ECI Optimization advances the policy toolkit of economic complexity by providing a mathematically grounded approach for strategic diversification.

In practice, policymakers should use ECI Optimization as a constrained, auditable shortlist of testable options rather than a blank-slate recommender. Each suggested item should be stress tested for windows of opportunity, institutional readiness, infrastructure, and market access. Also, these items should be linked to delivery instruments—skills, standards and certification, logistics improvements, or FDI facilitation. The suggested portfolio should be revisited on a fixed cadence with updated relatedness and PCI to track realized RCA and portfolio balance, retire weak candidates, and add new ones as conditions change.

In this context, ECI Optimization could be tailored for specific applications. For example, policymakers might care about raising value added, avoiding high-carbon or low unit-value products, leveraging domestic endowments, limiting commodity exposure, and securing local learning and regional balance. The method can be tailored to that logic by setting the objective (ECI, growth, inequality, or sustainability) and by shaping the admissible set with transparent constraints that preserve the tractability of the 0–1 program: exclude items below a unit-value floor or above a carbon-intensity ceiling; incentivize the use of a domestic-resource or supplier-base; or set up an investment fund to support the development of target activities.

Ultimately, ECI Optimization should motivate further advances in comprehensive methodologies for strategic diversification and real-world applications.

# Supplementary Material for: Optimizing Economic Complexity

**Abstract**

Efforts to apply economic complexity to identify diversification opportunities often rely on diagrams comparing the relatedness and complexity of products, technologies, or industries. Yet, the use of these diagrams, is not based on empirical or theoretical evidence supporting some notion of optimality. Here, we introduce an optimization-based framework that identifies diversification opportunities by minimizing a cost function capturing the constraints imposed by an economy's pattern of specialization. We show that the resulting portfolios often differ from those implied by relatedness–complexity diagrams, providing a target-oriented optimization layer to the economic complexity toolkit.

**Table of Contents**

## 1. Regression Model Performance in International Trade Data

In Figure S1, we present a heatmap of the average coefficient estimates, their standard errors, and p-values for entry models describing equation (2) in the main manuscript, estimated by using international trade data and varying the timeframe $\Delta t$ and stepping stone $\tau$. Figure S2 complements this by showing the average coefficient of determination for these models. We find that all coefficients exhibit a positive relationship with future specialization, as expected, and that, on average, they are statistically significant at the 0.01 level. Additionally, the models demonstrate an average coefficient of determination of 0.48, indicating a good fit. The coefficient of determination follows a gradient pattern, with higher values observed when $\tau$ is closer to $\Delta t$.

This pattern is primarily due to the autoregressive relationship between the RCA stepping stone $R_{cp}(t + \tau)$ and the dependent variable. As $\tau$ approaches $\Delta t$, the magnitude of the coefficient for $R_{cp}(t + \tau)$ increases, while the other coefficients correspondingly decrease.

Figures S3 and S4 illustrate the same heatmaps for the average coefficient estimates, standard errors, p-values, and coefficient of determination, but for the exit models. These models exhibit similar properties further underscoring the robustness of our approach. The only difference is that some of the coefficients lose statistical significance at certain thresholds.

Finally, in Figure S5 we display histograms of the distribution of the model coefficients when $\Delta t = 10$ and $\tau = 5$. We use this figure to demonstrate that the coefficients are tightly clustered around their mean values, with low variance, indicating consistency across different starting years. This clustering suggests that the model's estimates are similar regardless of the starting year.

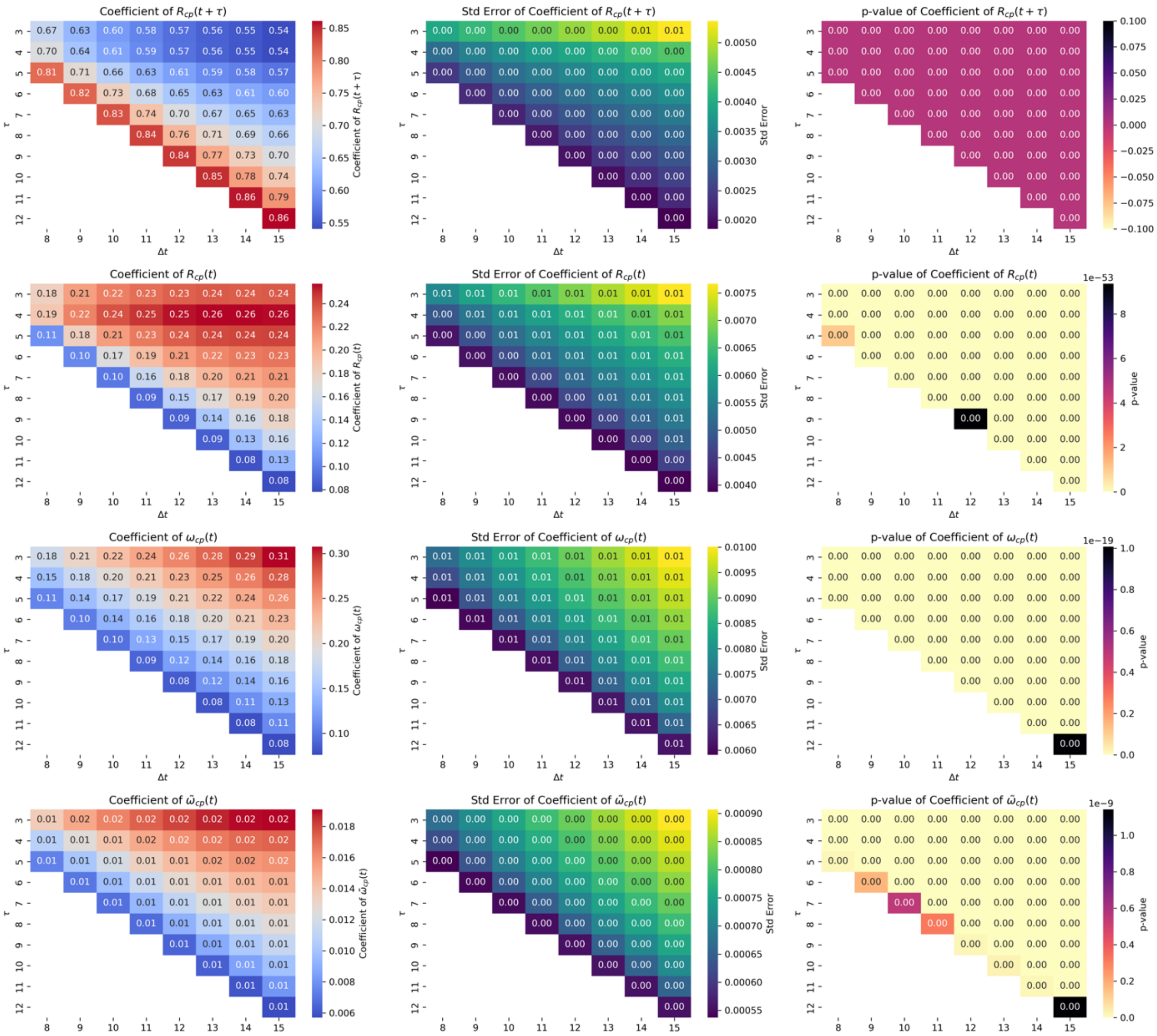

**Figure S1.** Heatmaps showing the average value of the coefficient estimates, their standard deviation, and their p-values as a function of timeframe $\Delta t$ and the steppingstone $\tau$, for entry regression models (defined as the subset of the data with $R_{cp}(t) < 1$) estimated using international trade data.

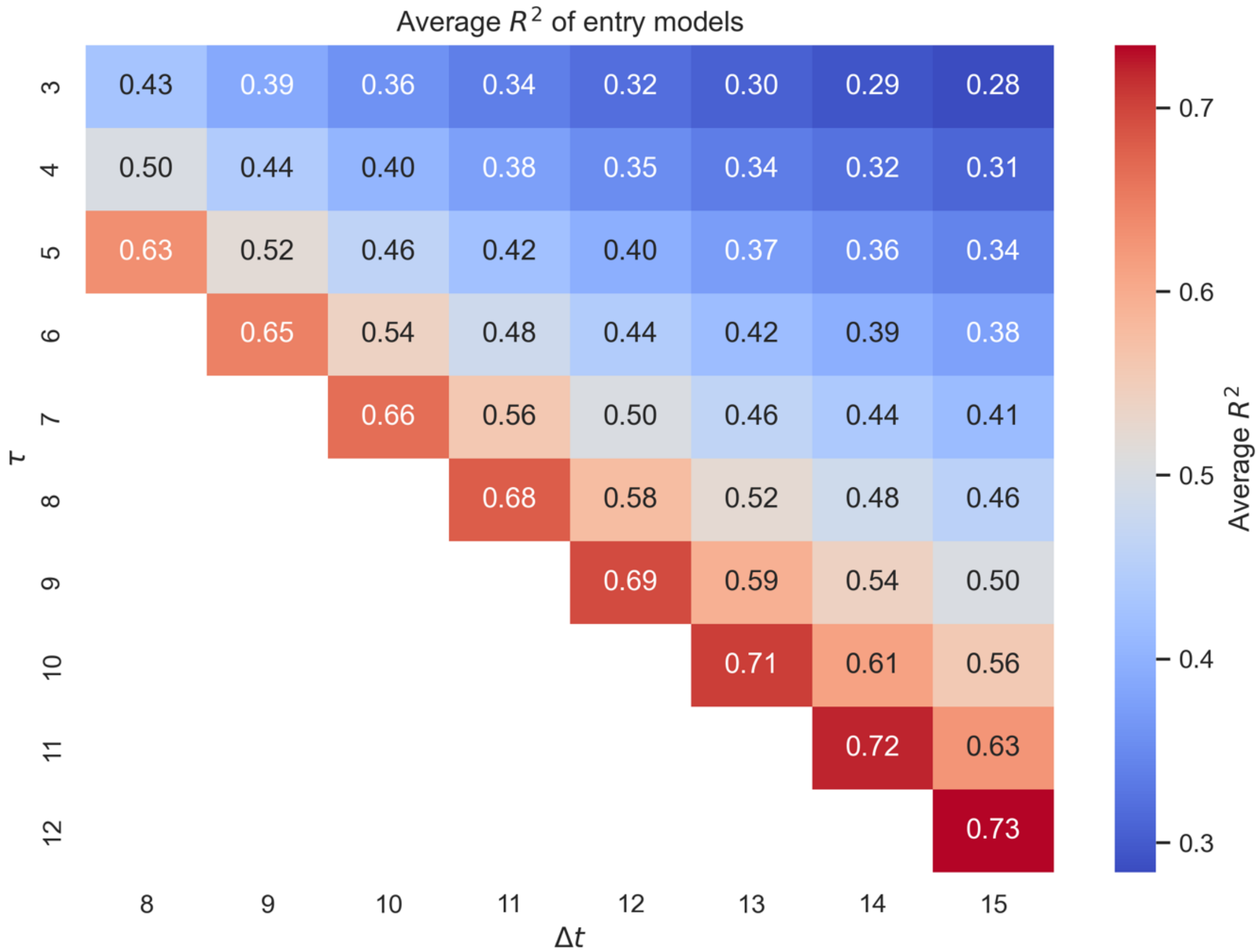

**Figure S2.** Heatmaps showing the coefficient of determination as a function of timeframe $\Delta t$ and the steppingstone $\tau$, for entry regression models (defined as the subset of the data with $R_{cp}(t) < 1$) estimated using international trade data.

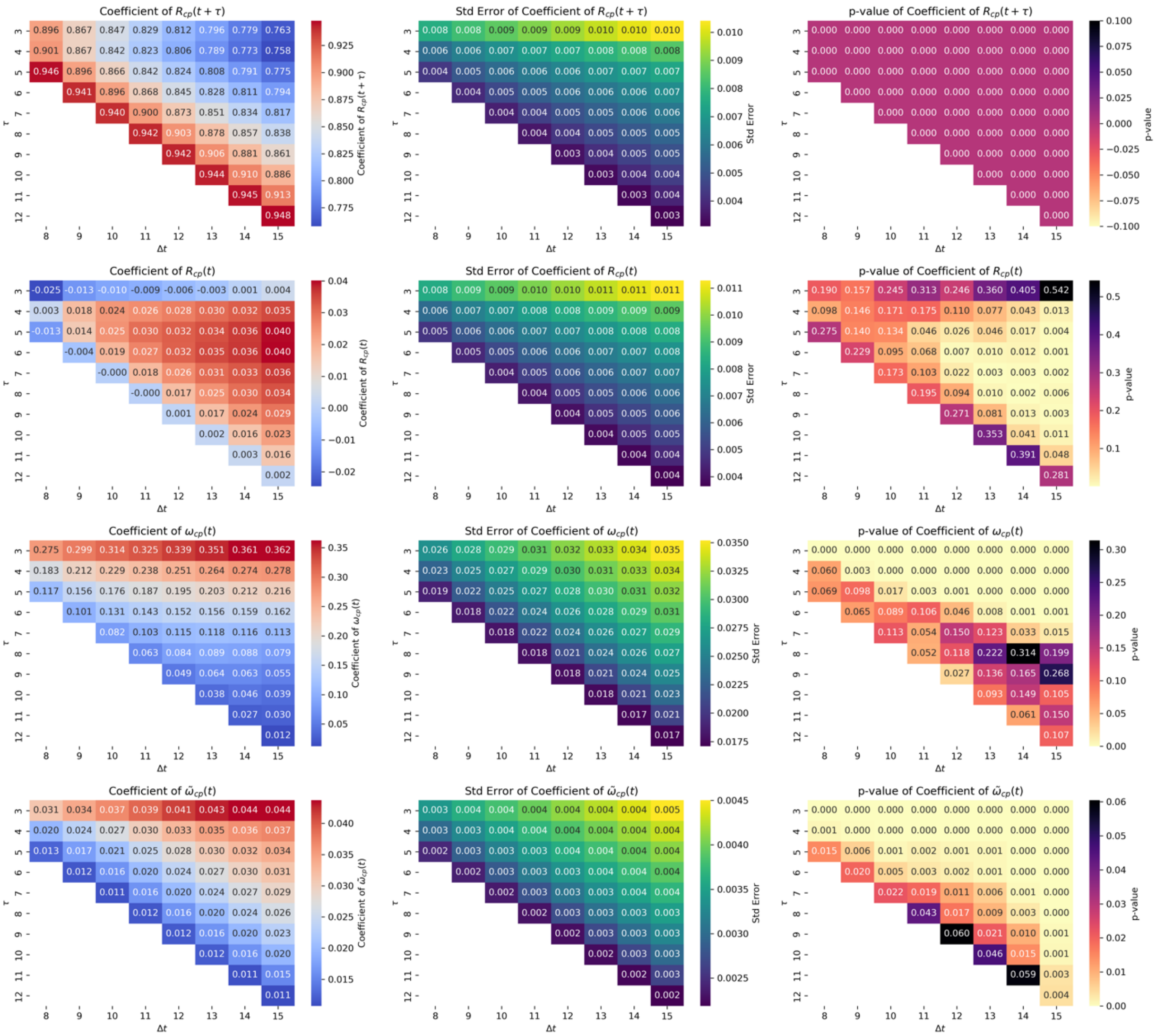

**Figure S3.** Heatmaps showing the average value of the coefficient estimates, their standard deviation, and their p-values as a function of timeframe $\Delta t$ and the steppingstone $\tau$, for exit regression models (defined as the subset of the data with $R_{cp}(t) \geq 1$) estimated using international trade data.

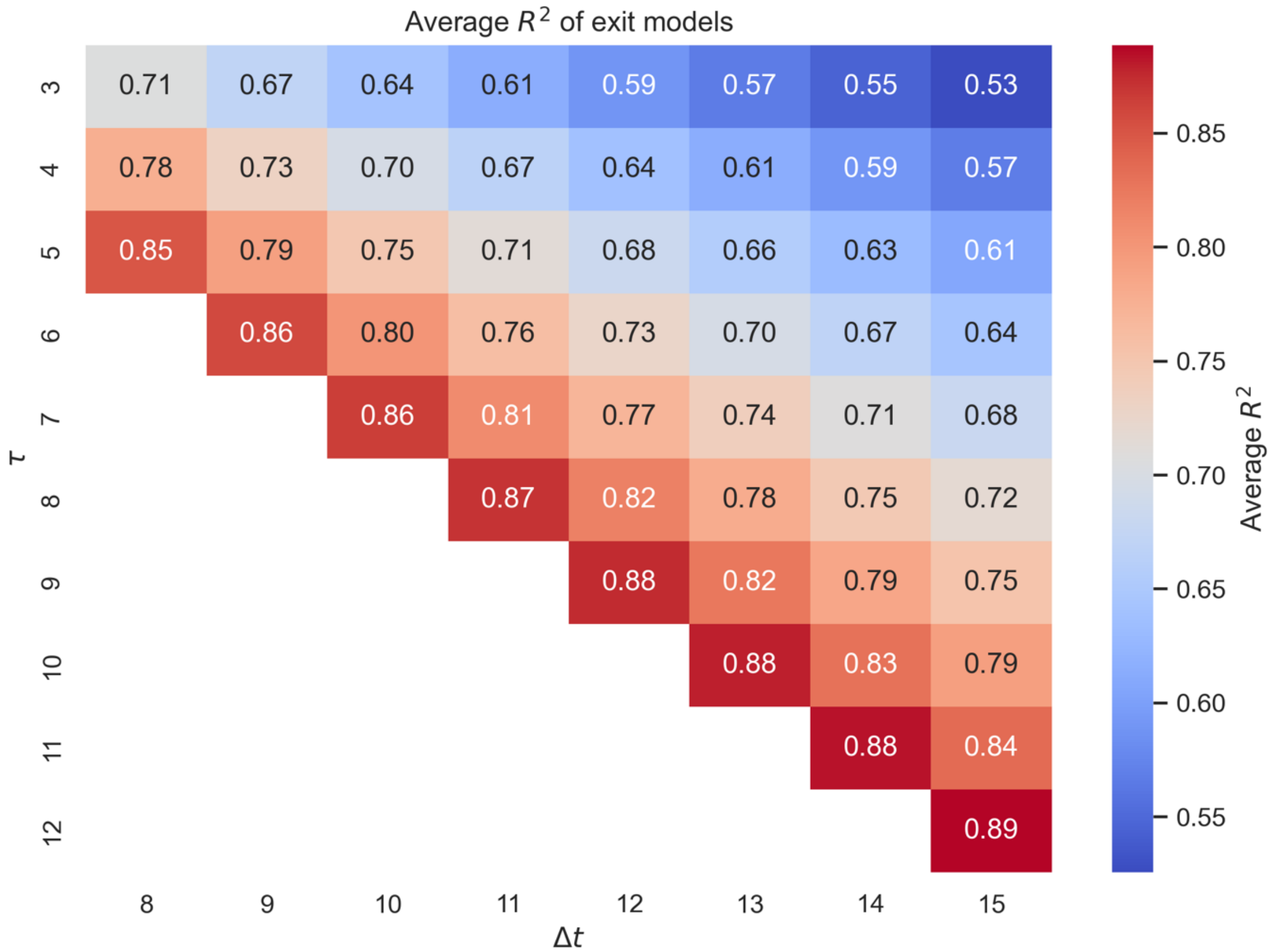

**Figure S4.** Heatmaps showing the coefficient of determination as a function of timeframe $\Delta t$ and the steppingstone $\tau$, for exit regression models (defined as the subset of the data with $R_{cp}(t) \geq 1$) estimated using international trade data.

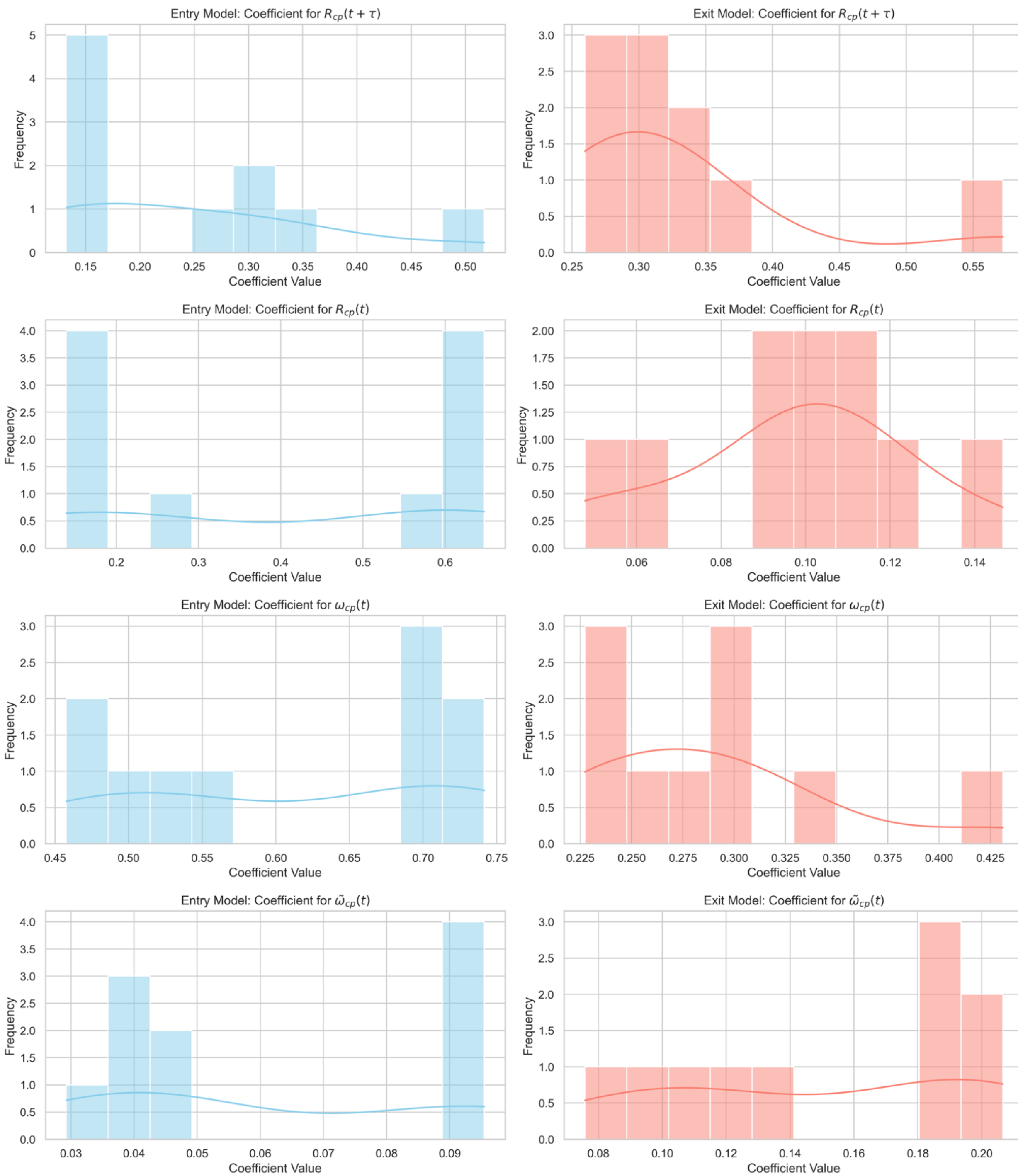

**Figure S5.** Histogram for the distribution of model coefficients for the models where $\Delta t = 10$ and $\tau = 5$ estimated using international trade data.

## 2. Regression Model Performance in The USA MSA Payroll Data

In Figure S6, we present a heatmap of the average coefficient estimates, their standard errors, and p-values for entry models describing equation (2), estimated by using the Usa MSA Payroll data and varying the timeframe $\Delta t$ and stepping stone $\tau$. Figure S7 complements this by showing the average coefficient of determination for these models. We find that all coefficients exhibit a positive relationship with future specialization, as expected, and that, most of the coefficients are statistically significant at the 0.01 level (only the relatedness coefficient needs a threshold of 0.1 in order to be significant when $\tau = 3$ and $\Delta t = 9$). These models demonstrate an average coefficient of determination of 0.25, indicating a more moderate fit when compared to the international trade data. Nevertheless, the coefficient of determination still follows a gradient pattern, with higher values observed when $\tau$ is closer to $\Delta t$.

Again, this pattern is primarily due to the autoregressive relationship between the RCA stepping stone $R_{cp}(t + \tau)$ and the dependent variable. As $\tau$ approaches $\Delta t$, the magnitude of the coefficient for $R_{cp}(t + \tau)$ increases, while the other coefficients correspondingly decrease.

Figures S8 and S9 illustrate the same heatmaps for the average coefficient estimates, standard errors, p-values, and coefficient of determination, but for the exit models. These models exhibit similar properties further underscoring the robustness across the MSA data.

Finally, in Figure S10 we display histograms of the distribution of the model coefficients when $\Delta t = 10$ and $\tau = 5$. We use this figure to demonstrate that the coefficients are tightly clustered around their mean values, with low variance, indicating consistency across different starting years. This clustering suggests that the model's estimates are similar regardless of the starting year.

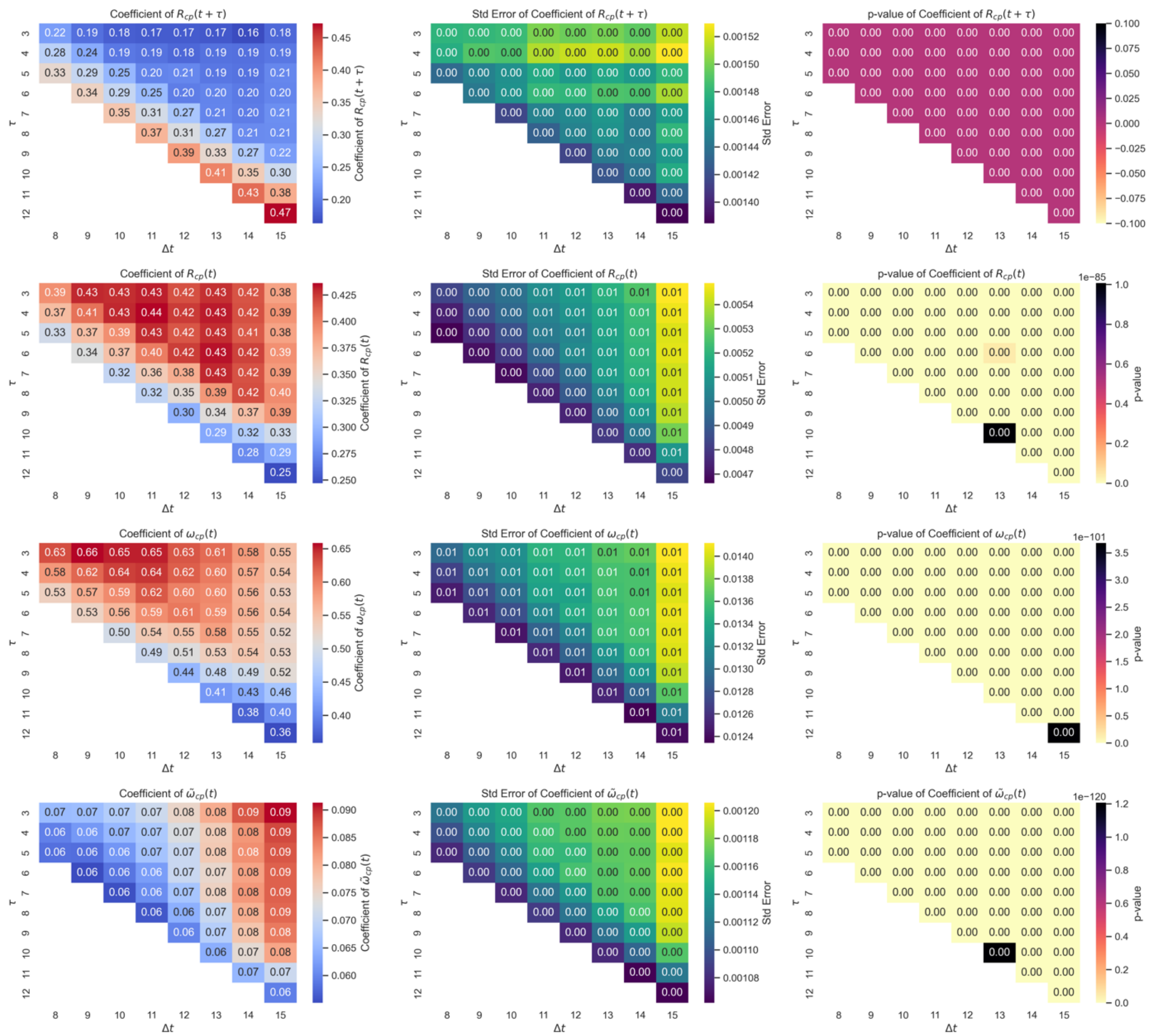

**Figure S6.** Heatmaps showing the average value of the coefficient estimates, their standard deviation, and their p-values as a function of timeframe $\Delta t$ and the steppingstone $\tau$, for entry regression models (defined as the subset of the data with $R_{cp}(t) < 1$) estimated using MSA payroll data.

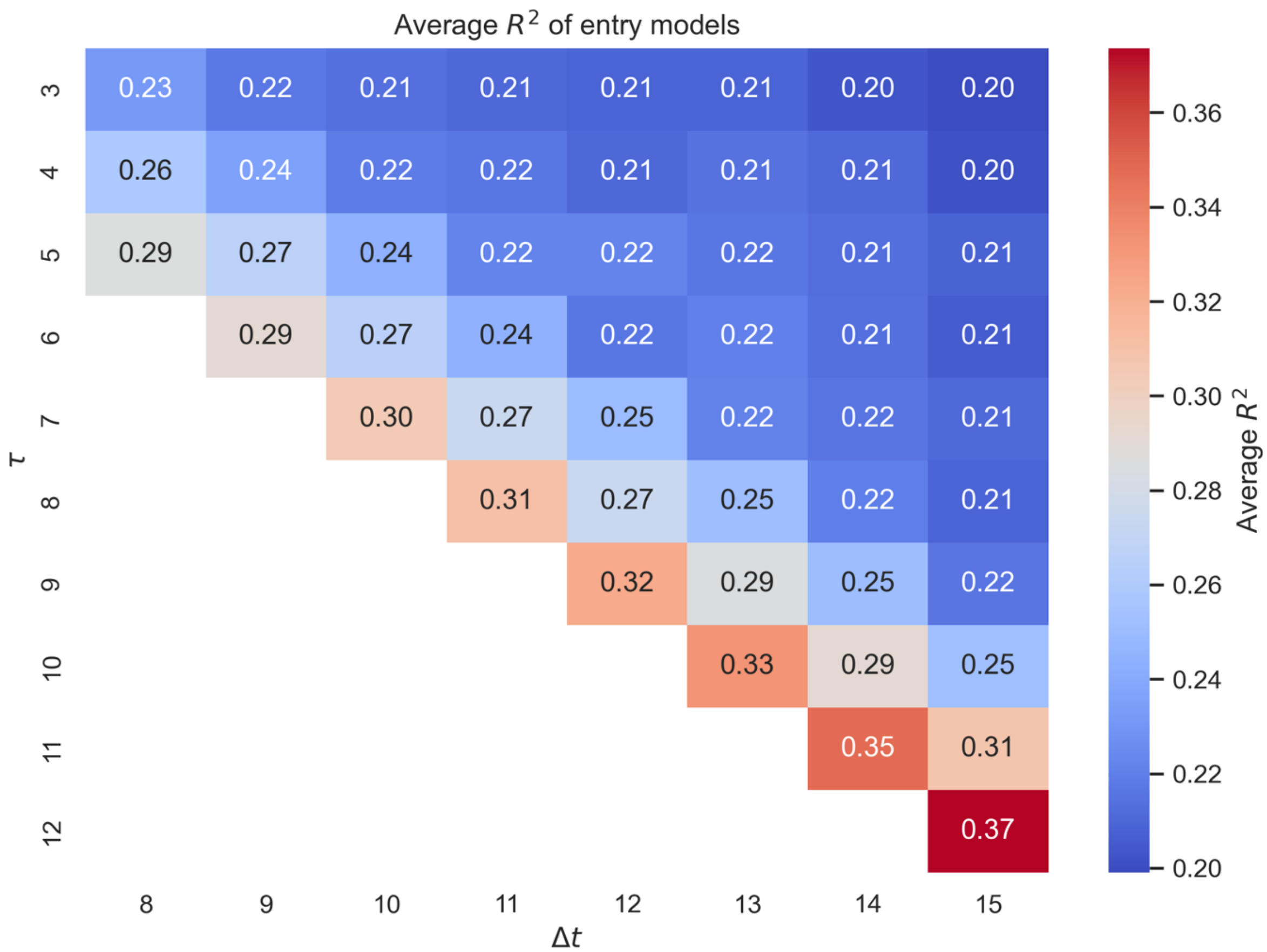

**Figure S7.** Heatmaps showing the coefficient of determination as a function of timeframe $\Delta t$ and the steppingstone $\tau$, for entry regression models (defined as the subset of the data with $R_{cp}(t) < 1$) estimated using MSA payroll data.

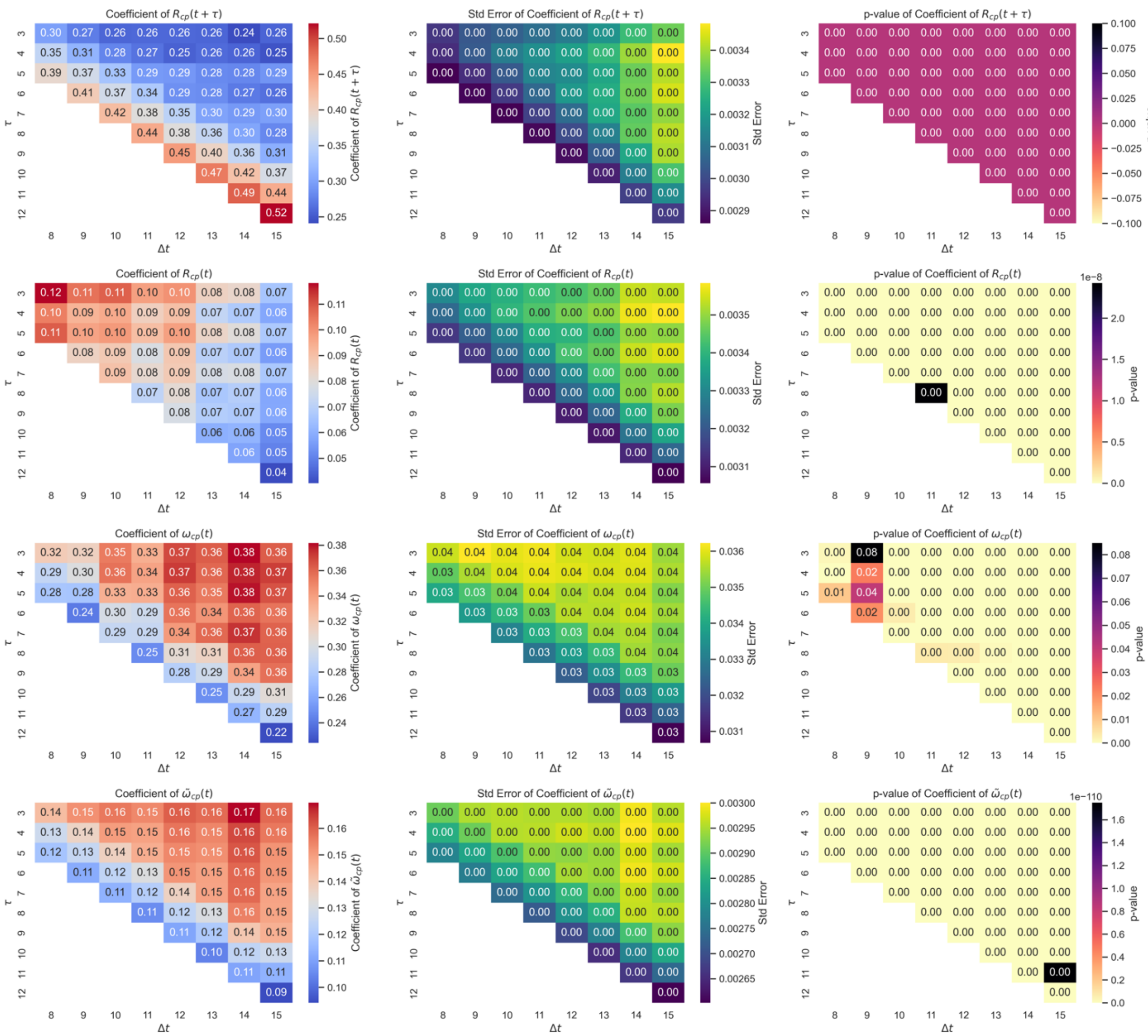

**Figure S8.** Heatmaps showing the average value of the coefficient estimates, their standard deviation, and their p-values as a function of timeframe $\Delta t$ and the steppingstone $\tau$, for exit regression models (defined as the subset of the data with $R_{cp}(t) > 1$) estimated using MSA payroll data.

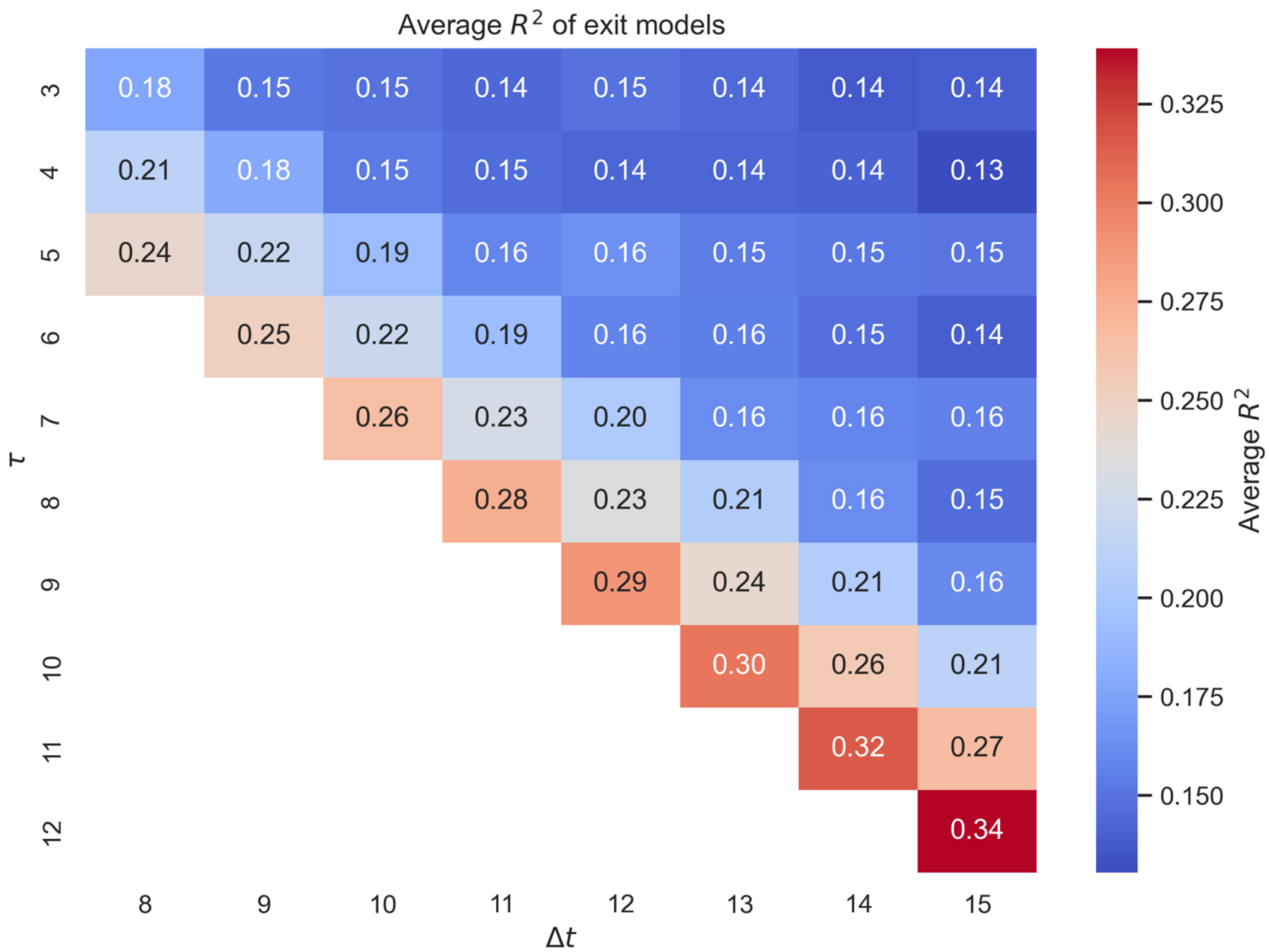

**Figure S9.** Heatmaps showing the coefficient of determination as a function of timeframe $\Delta t$ and the steppingstone $\tau$, for exit regression models (defined as the subset of the data with $R_{cp}(t) \geq 1$) estimated using MSA payroll data.

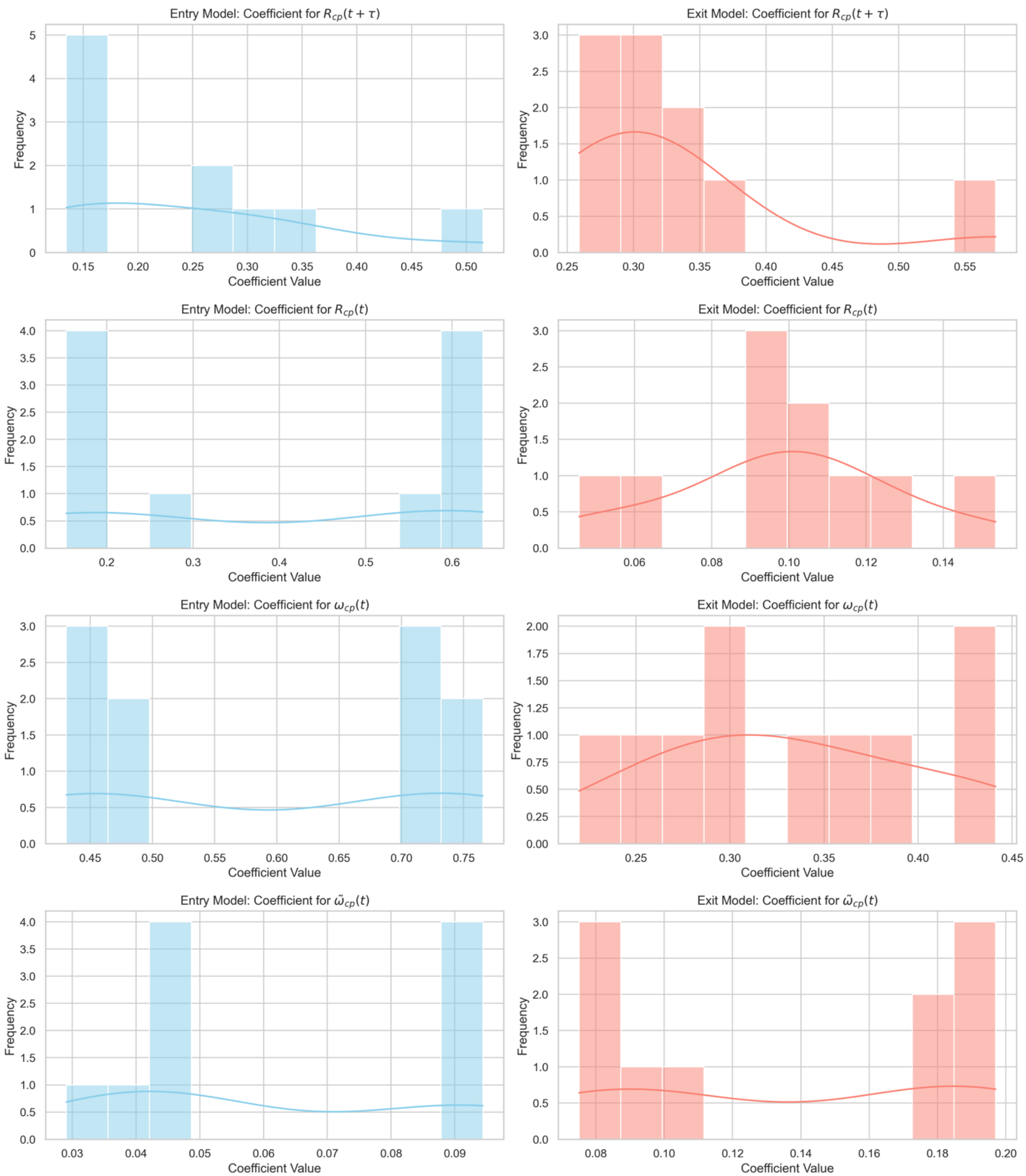

**Figure S10.** Histogram for the distribution of model coefficients for the models where $\Delta t = 10$ and $\tau = 5$ estimated using MSA payroll data.

## 3. Regression Model Performance in WIPO patent applications data

In Figure S11, we present a heatmap of the average coefficient estimates, their standard errors, and p-values for entry models describing equation (2) in the main manuscript, estimated by using international data on WIPO patent applications and varying the timeframe $\Delta t$ and stepping stone $\tau$. Figure S12 complements this by showing the average coefficient of determination for these models. We find that all coefficients exhibit a positive relationship with future specialization, as expected, and that, on average, they are statistically significant at the 0.01 level (the significance of the initial comparative advantage only decreases for larger timeframes). Additionally, the models demonstrate an average coefficient of determination of 0.12, indicating a low to moderate fit. The coefficient of determination follows a gradient pattern, with higher values observed when $\tau$ is closer to $\Delta t$.

This pattern is primarily due to the autoregressive relationship between the RCA stepping stone $R_{cp}(t + \tau)$ and the dependent variable. As $\tau$ approaches $\Delta t$, the magnitude of the coefficient for $R_{cp}(t + \tau)$ increases, while the other coefficients correspondingly decrease.

Figures S13 and S14 illustrate the same heatmaps for the average coefficient estimates, standard errors, p-values, and coefficient of determination, but for the exit models. These models exhibit similar properties further underscoring the robustness of our approach. The only difference is that some of the coefficients lose statistical significance at certain thresholds.

Finally, in Figure S15 we display histograms of the distribution of the model coefficients when $\Delta t = 10$ and $\tau = 5$. We use this figure to demonstrate that the coefficients are tightly clustered around their mean values, with low variance, indicating consistency across different starting years. This clustering suggests that the model's estimates are similar regardless of the starting year.

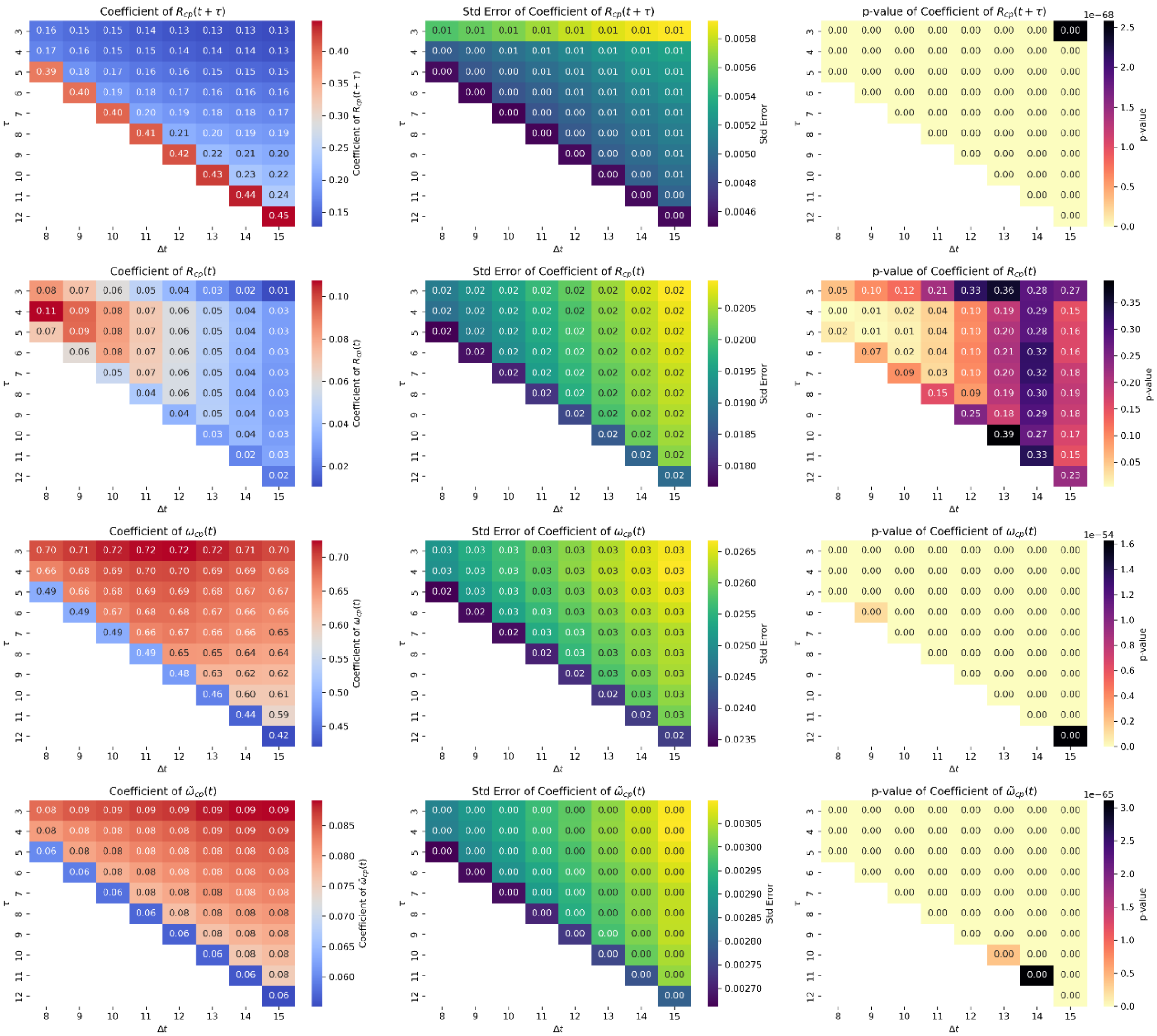

**Figure S11.** Heatmaps showing the average value of the coefficient estimates, their standard deviation, and their p-values as a function of timeframe $\Delta t$ and the steppingstone $\tau$, for entry regression models (defined as the subset of the data with $R_{cp}(t) < 1$) estimated using WIPO patent applications.

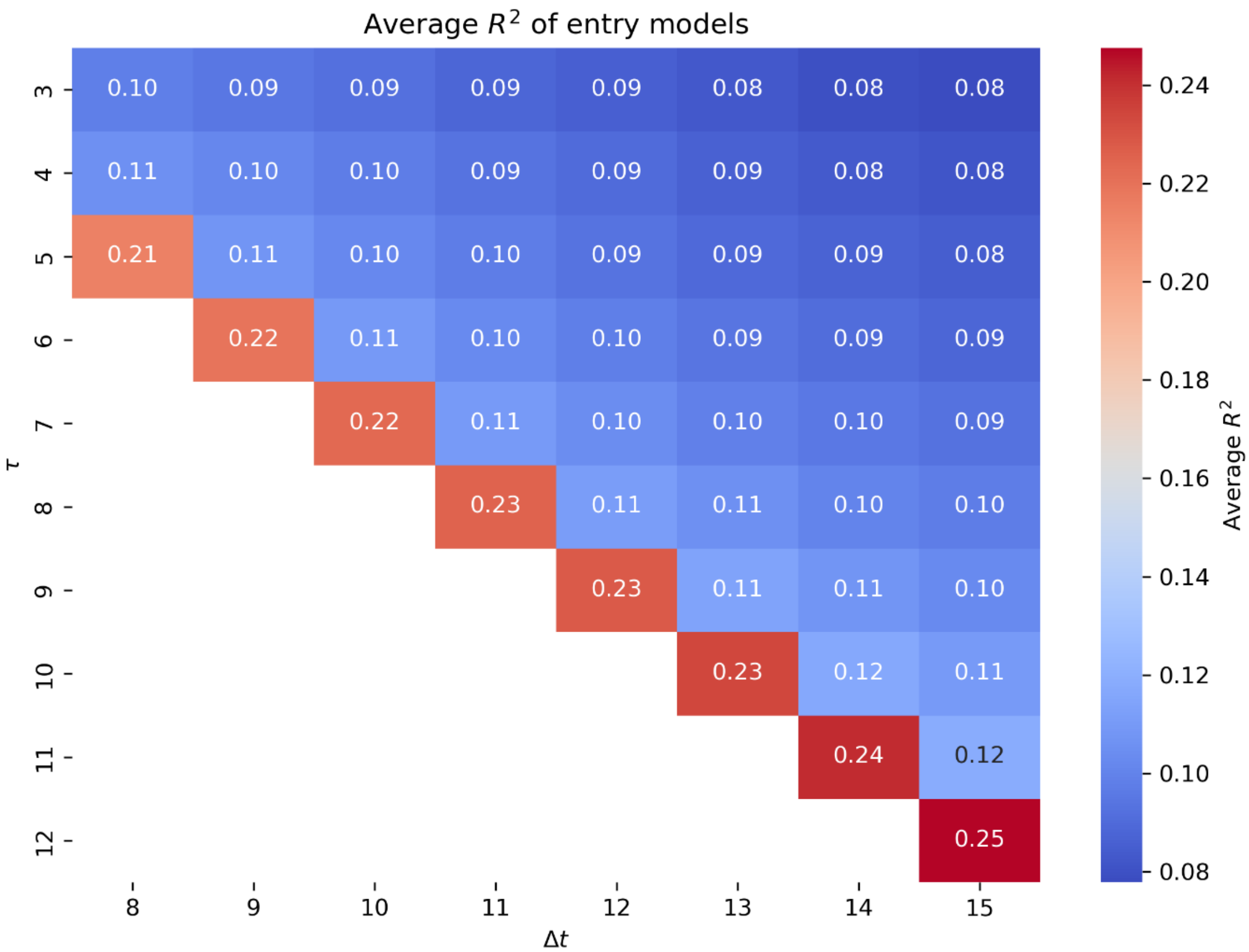

**Figure S12.** Heatmaps showing the coefficient of determination as a function of timeframe $\Delta t$ and the steppingstone $\tau$, for entry regression models (defined as the subset of the data with $R_{cp}(t) < 1$) estimated using WIPO patent applications.

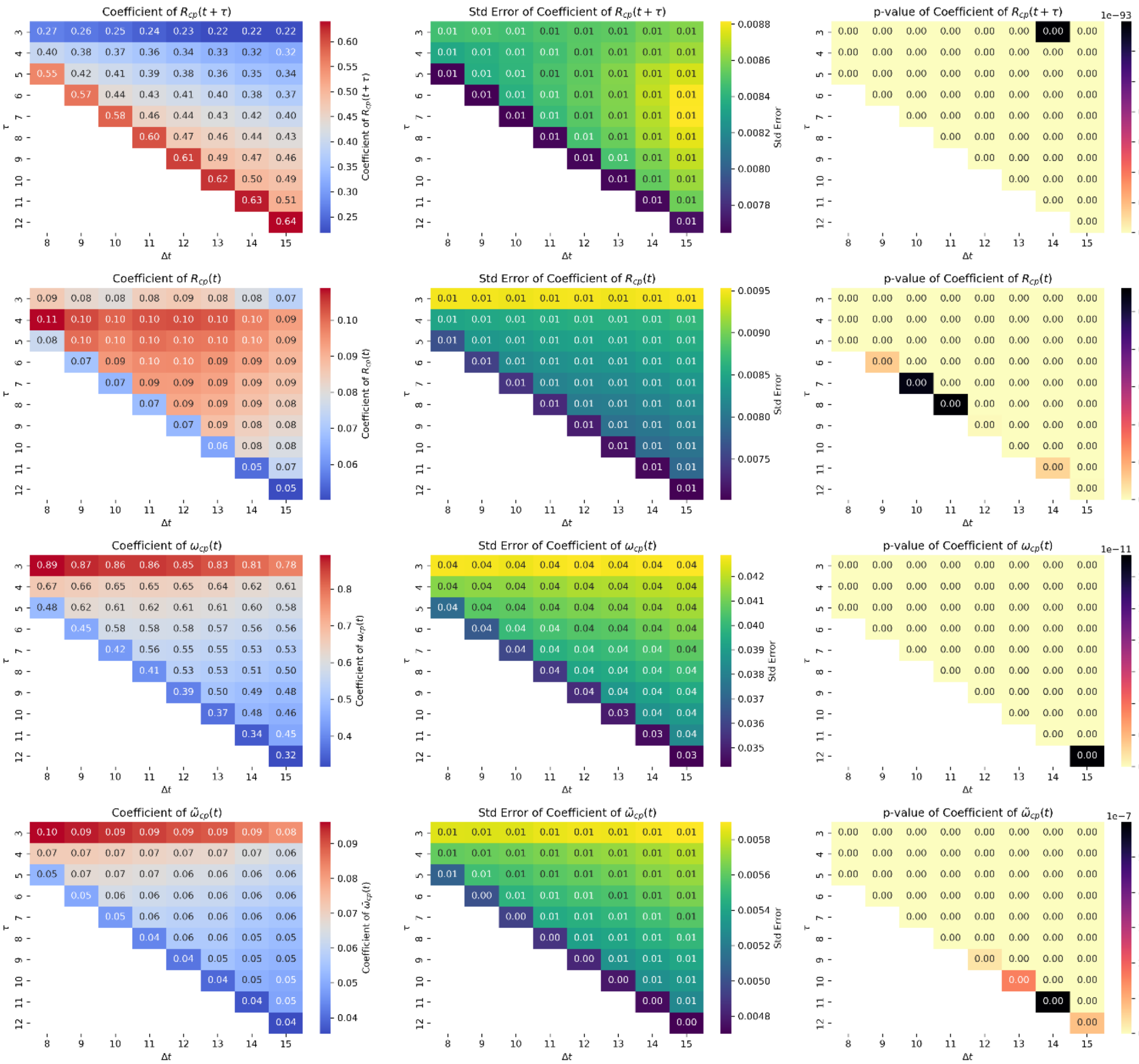

**Figure S13.** Heatmaps showing the average value of the coefficient estimates, their standard deviation, and their p-values as a function of timeframe $\Delta t$ and the steppingstone $\tau$, for exit regression models (defined as the subset of the data with $R_{cp}(t) \geq 1$) estimated using WIPO patent applications.

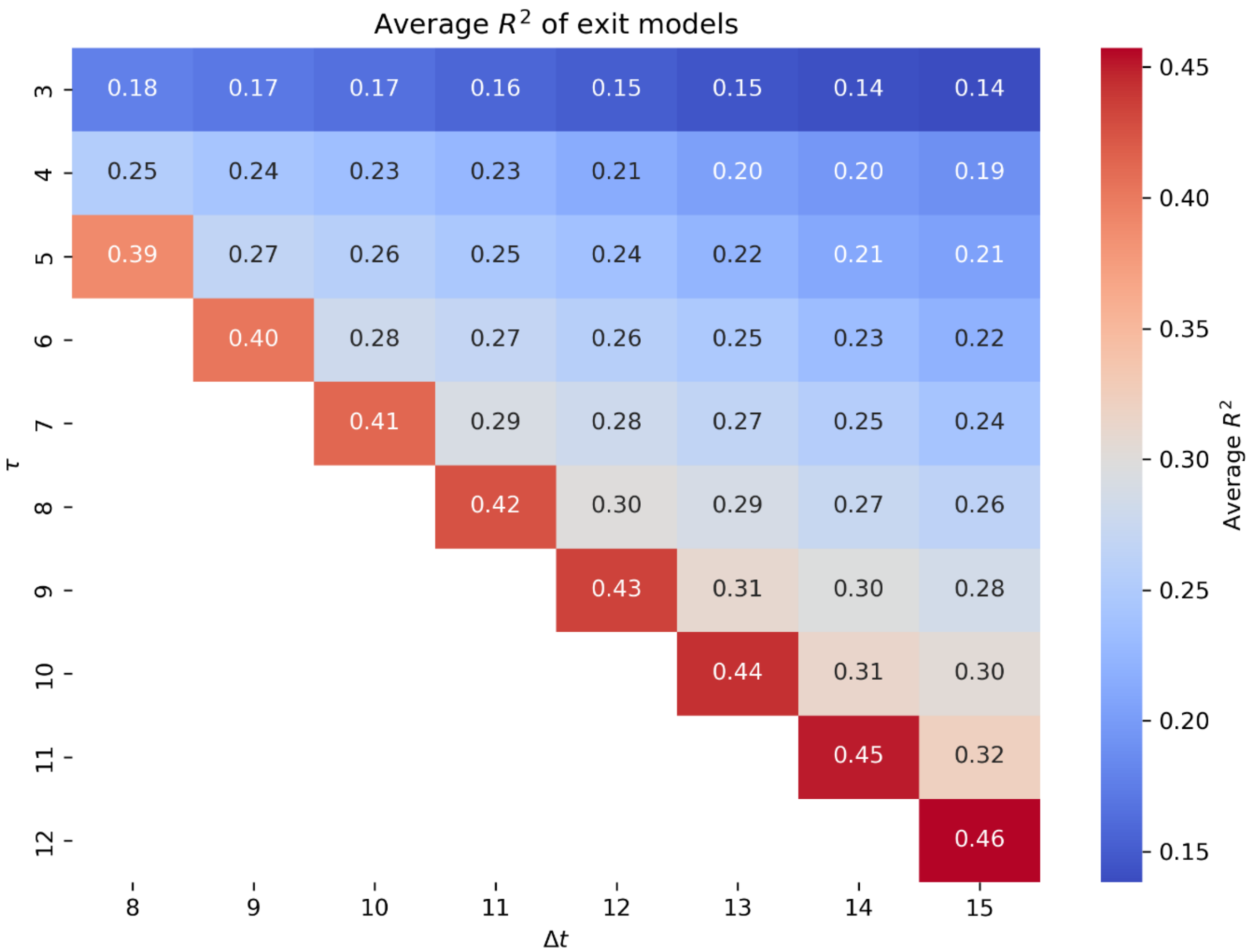

**Figure S14.** Heatmaps showing the coefficient of determination as a function of timeframe $\Delta t$ and the steppingstone $\tau$, for exit regression models (defined as the subset of the data with $R_{cp}(t) \geq 1$) estimated using WIPO patent applications.

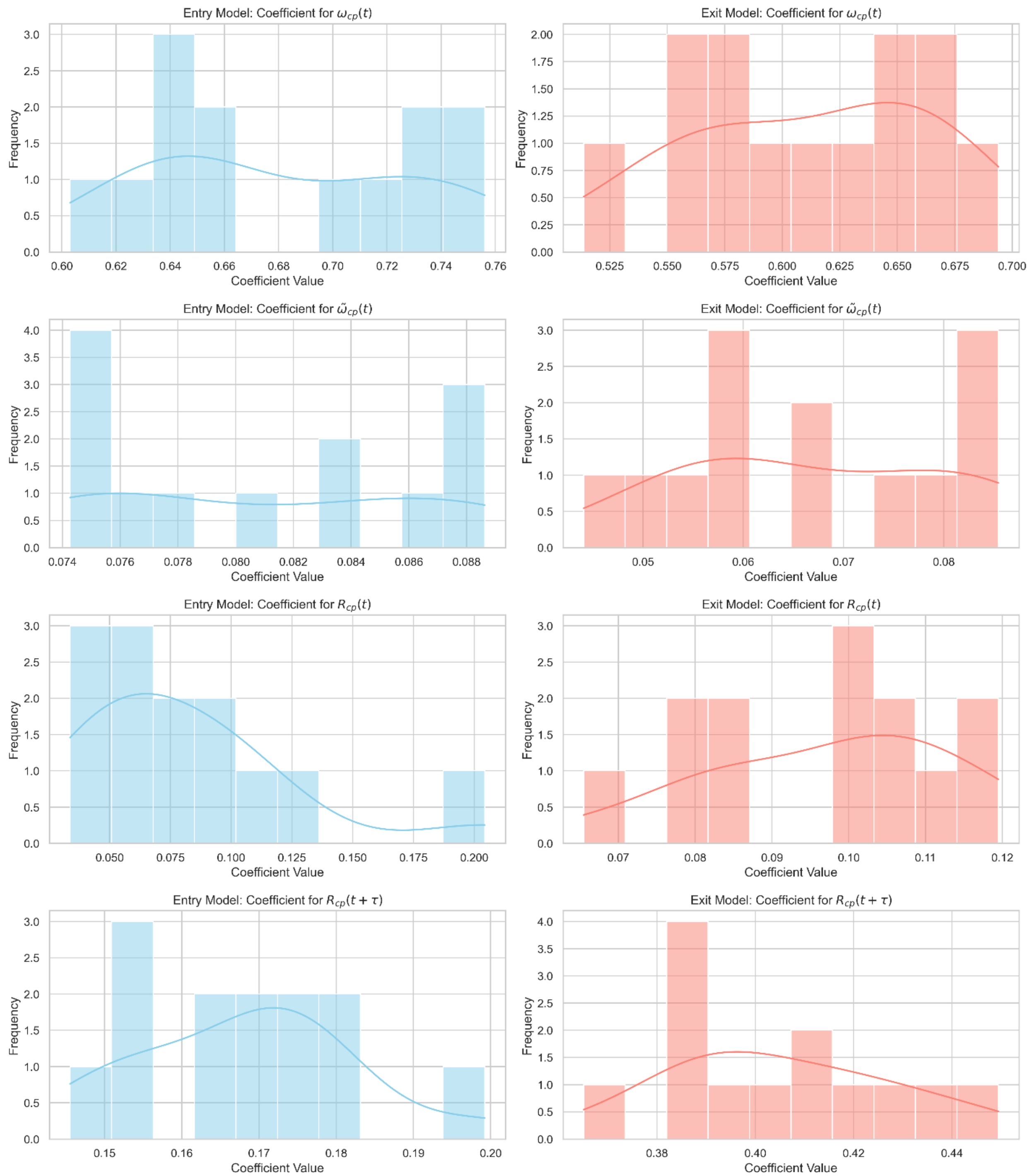

**Figure S15.** Histogram for the distribution of model coefficients for the models where $\Delta t = 10$ and $\tau = 5$ estimated using WIPO patent applications.

## 4. Properties of ECI Optimization in the USA MSA Payroll Data

In Figure S11, we repeat this analysis described in the section "Properties of ECI Optimzation" using MSA payroll data. Again, we find that ECI optimization suggests activities that are more aligned with the MSA's current specialization (Figure S16a). But unlike the case of international trade data, we do not observe the U-shaped relationship between ECI and average relative relatedness (Figure S16b). Despite this, ECI optimization still provides suggestions that are of slightly lower relatedness compared to the benchmark model. Additionally, the number of new activities and the added volume suggested by the ECI optimization method are, once again, lower than the benchmark model (Figures S16c and S16d), indicating a balanced approach in terms of both diversification and investment.

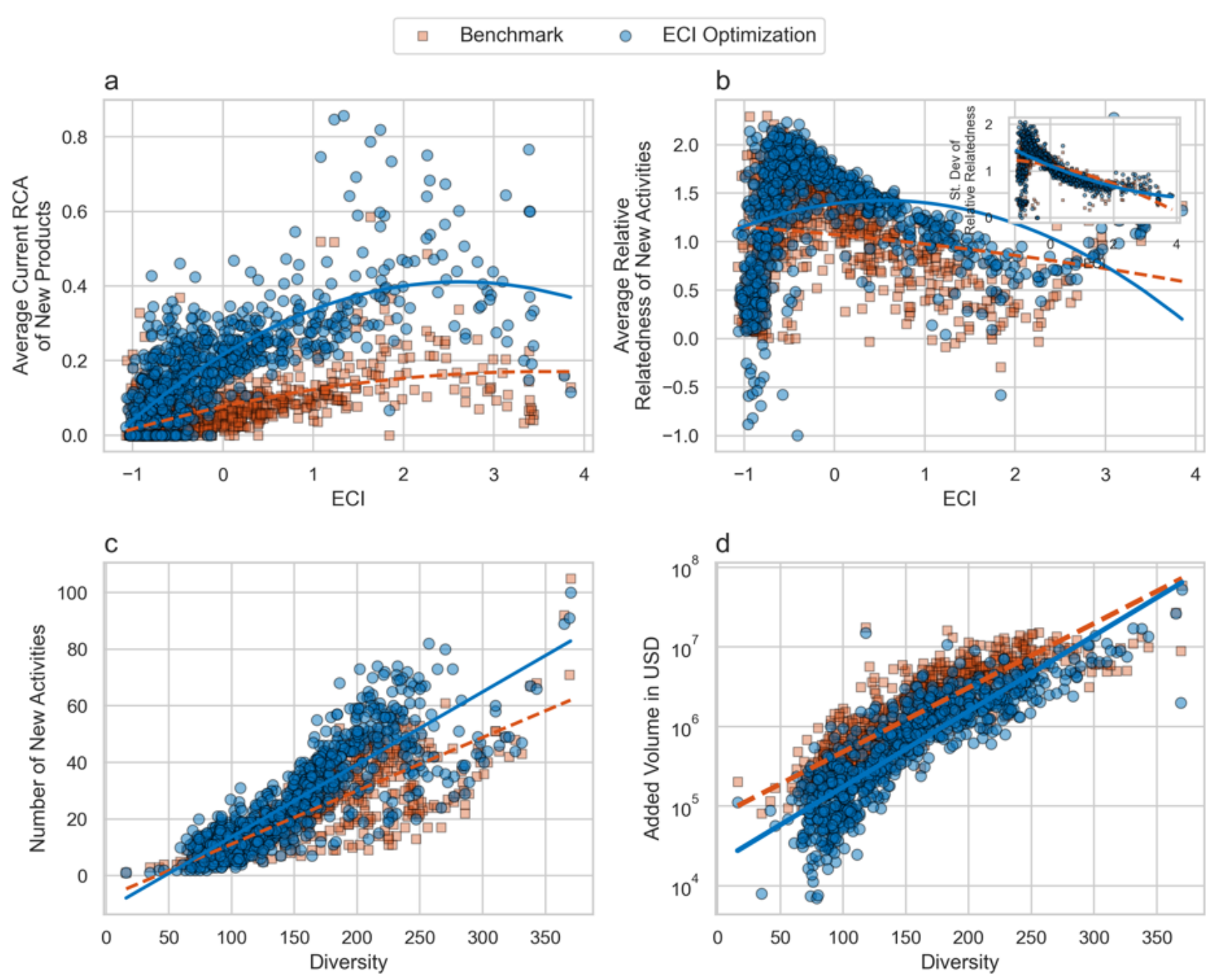

**Figure S16. Properties of ECI Optimization in United States MSA Data.** **a)** Average RCA (in 2022) of the activities suggested by ECI optimization and the benchmark model as a function of an MSA's ECI in 2022. **b)** Average relative relatedness of the suggested activities as a function of an MSA's ECI in 2022. The inset plot shows the variance of the relative relatedness of the suggested products. **c)** The number of suggested activities as a function of the initial diversity (number of activities in which the MSA had comparative advantage in 2022). **d)** Estimated added payroll volume to gain comparative advantage in the suggested activities as a function of the initial diversity. **a-d)** For each MSA we assume an increase of 0.1 of the ECI value in 2022. Each scatter chart has a quadratic fit line.

## 5. Properties of ECI Optimization in the WIPO Patent Applications Data

In Figure S17, we repeat this analysis described in the section "Properties of ECI Optimzation" using WIPO patent applications data. Again, we find that ECI optimization suggests activities that are more aligned with the country's current specialization (Figure S17a). But unlike the case of international trade data, we do not observe as smooth U-shaped relationship between ECI and average relative relatedness (Figure S17b). Despite this, ECI optimization still provides suggestions that are of slightly lower relatedness compared to the benchmark model. Additionally, the number of new activities and the added volume suggested by the ECI optimization method are, only slightly lower than the benchmark model (Figures S17c and S17d), indicating a balanced approach in terms of both diversification and investment.

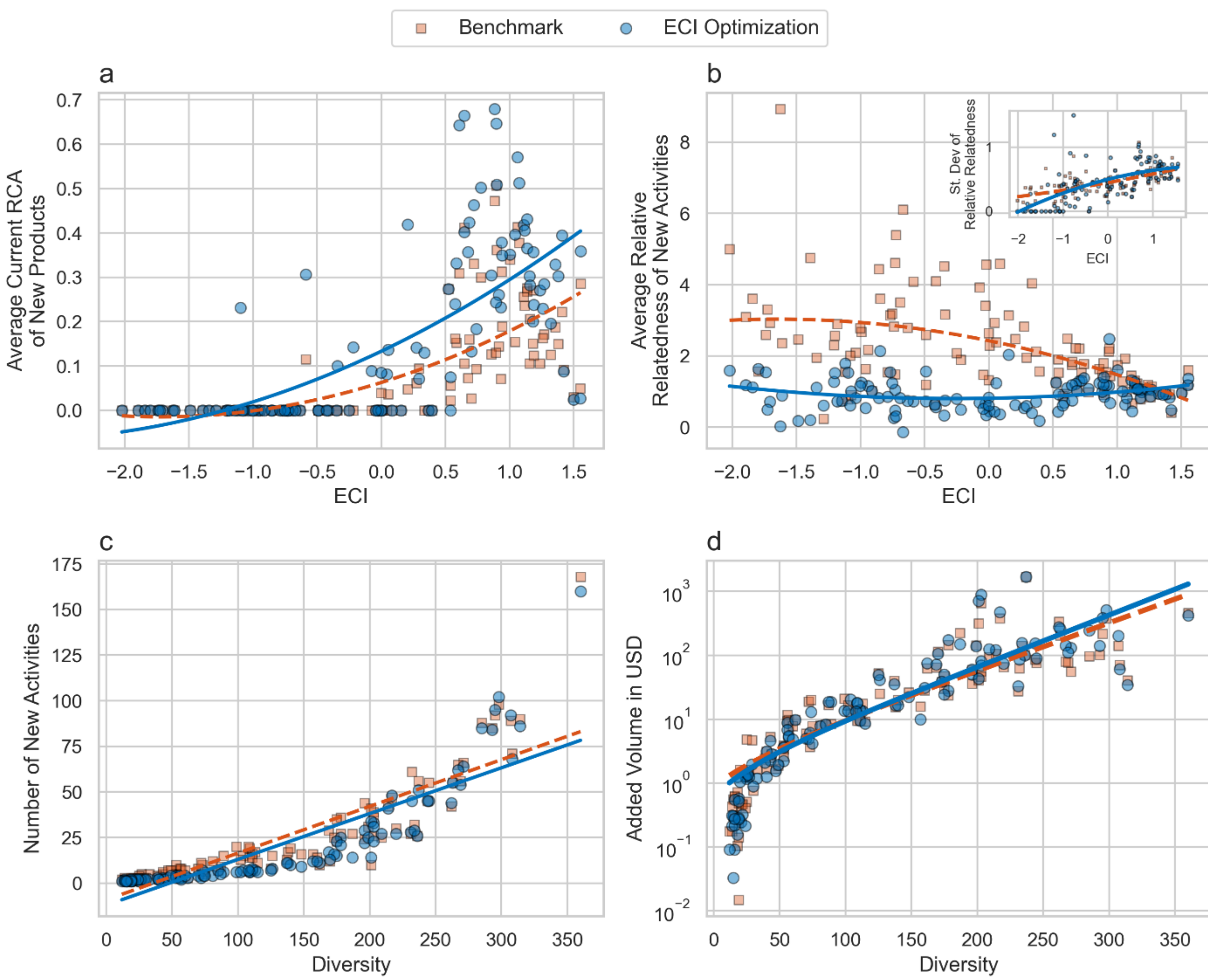

**Figure S17. Properties of ECI Optimization in WIPO Patent applications data. a)** Average RCA (in 2021) of the activities suggested by ECI optimization and the benchmark model as a function of a country's ECI in 2021. **b)** Average relative relatedness of the suggested activities as a function of a country's ECI in 2021. The inset plot shows the variance of the relative relatedness of the suggested products. **c)** The number of suggested activities as a function of the initial diversity (number of activities in which the country had comparative advantage in 2021). **d)** Estimated added payroll volume to gain comparative advantage in the suggested activities as a function of the initial diversity. **a-d)** For each country we assume an increase of 0.1 of the ECI value in 2021. Each scatter chart has a quadratic fit line.

## 6. Economic Growth Regression Model Results

In Table S1 we show the results of our economic growth model regressions used in the framework to predict economic structures given a target growth rate.

In these regressions, the dependent variable is the annualized 10-year growth rate of GDP per capita (in PPP constant 2021 USD) of a country (we consider two periods 1999–2009 and 2009–2019). We use two additional explanatory variables. First, we use a z-score normalized value for the log of initial GDP per capita (normalized across our sample of countries for each year), capturing Solow's idea of economic convergence. This z-score transformation helps us account for the non-stationary nature of GDP per capita, and hence provide a consistent prediction. Second, we use the interaction of ECI with the z-score of the initial log of GDP per capita, capturing the idea that the contribution of economic complexity to future economic growth depends on the current level of income.[9] We also use period fixed effects in order to account for any omitted variables that vary over the two decades and may impact economic growth.

In column (1) of Table S1 we show the results of a baseline model incorporating only the Solow term, in column (2) we introduce ECI to the equation, and in model (3) we combine ECI and gdp per capita to predict economic growth. In each case, all of the variables are statistically significant. More importantly ECI significantly upgrades the baselines model predictive power. Namely, the R-squared grows from 0.09 in the baseline model to 0.19 when we use just ECI as a predictor, and to 0.23 when we combine ECI and GDP per capita. This provides a statistical validity for the growth model used in the main manuscript. Nevertheless, we emphasize that more complex models incorporating multiple explanatory variables could potentially increase the explanatory power (but might decrease the available data).

**Table S1. Economic Growth Regression Results.**

| | *Dependent variable: Annualized growth of GDP per capita (in PPP constant 2021 USD) (1999-2009, 2009-2019)* | | |
|---|---|---|---|
| | (1) | (2) | (3) |
| ECI (trade) | | $0.868^{***}$ | $1.073^{***}$ |
| | | (0.199) | (0.199) |
| ECI (trade) x Log of initial GDP per capita | | | $-0.461^{***}$ |
| | | | (0.150) |
| Log of initial GDP per capita | $-0.714^{***}$ | $-1.392^{***}$ | $-1.536^{***}$ |
| | (0.163) | (0.235) | (0.229) |
| Observations | 172 | 172 | 172 |
| $R^2$ | 0.092 | 0.192 | 0.229 |
| Adjusted $R^2$ | 0.086 | 0.182 | 0.215 |
| Residual Std. Error | 1.954 (df=170) | 1.848 (df=169) | 1.811 (df=168) |
| F Statistic | $19.233^{***}$ (df=1; 170) | $17.630^{***}$ (df=2; 169) | $15.409^{***}$ (df=3; 168) |

Note: $^{*}p<0.1$; $^{**}p<0.05$; $^{***}p<0.01$. Robust standard errors in parentheses.